\documentclass[acmsmall]{acmart}

\usepackage{multirow}
\usepackage{booktabs}
\usepackage{xcolor}
\usepackage{hhline}
\usepackage{adjustbox}
\usepackage{makecell}
\usepackage{cleveref}
\usepackage{longtable}
\usepackage{enumitem}
\usepackage{changepage} % put in preamble
\usepackage{verbatim}
\usepackage[frozencache,cachedir=.]{minted}
\usepackage{soul}
\usepackage[most]{tcolorbox}  
\usepackage[frozencache,cachedir=.]{minted}
\usepackage{soul}
\usepackage[normalem]{ulem}  
\usepackage{fontawesome5} 
\usepackage{colortbl} 
\usepackage{float}
\usepackage{placeins}
\usepackage[edges]{forest}
\usepackage[frozencache,cachedir=.]{minted}
\usepackage{svg}
\usepackage{pgfplots}
\pgfplotsset{compat=1.18}
\pgfplotsset{compat=1.17}
\usepackage{tabularx}
\usepgfplotslibrary{statistics}
\pgfplotsset{compat=newest}
\usepackage{multirow}
\usetikzlibrary{patterns}
\usepgfplotslibrary{groupplots}
\usepackage{subcaption}
\setcopyright{cc}
\setcctype{by}
\acmDOI{10.1145/3832285}
\acmYear{2026}
\acmJournal{PACMSE}
\acmVolume{3}
\acmNumber{ISSTA}
\acmArticle{ISSTA194}
\acmMonth{10}
\acmSubmissionID{issta26main-p2374-p}
\makeatletter
\patchcmd{\@mkbibcitation}{\par\medskip\small\noindent}{\par\medskip\footnotesize\noindent}{}{}
\makeatother
\begin{document}
\received{2026-01-30}
\received[accepted]{2026-04-16}

% \title{Understanding APR Agents Through the Lens of Traceability: An Empirical Study}
\title{Understanding Automated Program Repair Agents through the Lens of Traceability: An Empirical Study}

\author{Ira Ceka}
\authornote{Equal contribution as first authors.}
\orcid{0000-0003-4697-5586}
\affiliation{%
  \institution{Columbia University}
  \city{New York}
  \country{USA}
}
\email{iceka@cs.columbia.edu}

\author{Hailie Mitchell}
\authornotemark[1]
\orcid{0009-0004-8010-5028}
\affiliation{%
  \institution{Columbia University}
  \city{New York}
  \country{USA}
}
\email{hailie.m@cs.columbia.edu}

\author{Saurabh Pujar}
\authornote{Corresponding author.}
\orcid{0000-0002-9772-3162}
\affiliation{%
  \institution{IBM Research}
  \city{Yorktown}
  \country{USA}
}
\email{saurabh.pujar@ibm.com}

\author{Luca Buratti}
\orcid{0009-0007-1468-9995}
\affiliation{%
  \institution{IBM Research}
  \city{Yorktown}
  \country{USA}
}
\email{luca.buratti1@ibm.com}

\author{Shyam Ramji}
\orcid{0009-0006-1968-8740}
\affiliation{%
  \institution{IBM Research}
  \city{Yorktown}
  \country{USA}
}
\email{ramji@us.ibm.com}

\author{Junfeng Yang}
\orcid{0009-0000-2277-6545}
\affiliation{%
  \institution{Columbia University}
  \city{New York}
  \country{USA}
}
\email{junfeng@cs.columbia.edu}

\author{Gail Kaiser}
\orcid{0000-0002-8791-1178}
\affiliation{%
  \institution{Columbia University}
  \city{New York}
  \country{USA}
}
\email{kaiser@cs.columbia.edu}

\author{Baishakhi Ray}
\orcid{0000-0003-3406-5235}
\affiliation{%
  \institution{Columbia University}
  \city{New York}
  \country{USA}
}
\email{rayb@cs.columbia.edu}

\renewcommand{\shortauthors}{I. Ceka, H. Mitchell, S. Pujar, L. Buratti, S. Ramji, J. Yang, G. Kaiser, B. Ray}

\newtcolorbox{findingbox}{
  enhanced,
  breakable,                
  colback  = blue!10,       
  colframe = blue!10!black,
  coltitle = black,         
  fonttitle=\bfseries,      
  top    = 1mm,
  bottom = 1mm,
  left   = 1.5mm,
  right  = 1.5mm,
  boxrule = 0.6pt,           
  arc      = 2mm,            
  width=\linewidth           
}

\newtcblisting{motivationbox}[1][]{
  listing only,
  %breakable,
  colback=codebg,
  colframe=codeframe,
  boxrule=0.5pt,
  arc=2pt,
  left=2pt, right=2pt,
  top=2pt, bottom=2pt,
  enhanced,
  width=\linewidth,
  before skip=3pt,
  after skip=3pt,
  listing options={
    language=MyBash,
    basicstyle=\verytinymono,
    breaklines=true,
    breakatwhitespace=false,
    columns=flexible
  },
  #1
}

\definecolor{codeframe}{gray}{0.60}

\newcommand{\verytinymono}{\fontsize{7.5}{7}\ttfamily\selectfont}

\lstdefinelanguage{MyBash}{
  keywords={set,source,conda,cd,git,python},
  keywordstyle=\color{blue!75!black}\bfseries,
  comment=[l]{\#}, commentstyle=\color{green!50!black},
  stringstyle=\color{red!75!black},
  morestring=[b]', morestring=[b]"
}

\lstset{
  language=MyBash,
  basicstyle=\verytinymono,
  breaklines=true,
  breakatwhitespace=true,
  showstringspaces=false,
  tabsize=2
}

\definecolor{codebg}{gray}{0.95} 

\newcommand{\e}{{Easy}\xspace}
\newcommand{\m}{{Moderate}\xspace}
\newcommand{\h}{{Hard}\xspace}
\newcommand{\vh}{{Very Hard}\xspace}


\begin{abstract}
Automated Program Repair (APR) agents leverage large language models (LLMs) to autonomously diagnose and patch software bugs using planning, reasoning, and tools. Although these agents show strong performance on leaderboards such as SWE-bench, little is understood about how they take actions, where they fail, and how their behavior compares to human developers. In this paper, we present the first systematic analysis of these limitations using 5 state-of-the-art APR agents. We trace the full decision-making pipelines of the 5 APR agents across 500 real-world repair tasks, from issue description to patch validation. 

Our study reveals that, while agents excel at simple fixes, they struggle with logic-intensive bugs, often generating verbose, overfitted patches that pass existing test suites without solving the root cause. Test generation and regression test selection remain major bottlenecks, as agents fail to reproduce issues or run relevant regression tests. Moreover, many agents operate with primitive tooling (e.g. bash scripts) and do not have access to debuggers or program analysis tools. These findings highlight key limitations of current APR systems and motivate several directions for next-generation APR design, including but not limited to: (1) a shift-left approach emphasizing early, high-quality test generation and validation to reduce spurious fixes and improve semantic correctness; (2) richer, more integrated tool ecosystems; (3) diversified agent architectures that combine complementary strengths; and (4) benchmarks that prioritize semantic repair quality and test-generation fidelity over surface-level success metrics.

\end{abstract}


\maketitle

% copied from another paper

\definecolor{MyColor}{RGB}{50, 100, 250}
\definecolor{Orange}{RGB}{244, 101, 66}
\definecolor{Red}{RGB}{255, 0, 0}
\definecolor{Green}{RGB}{0, 255, 0}
\definecolor{Blue}{RGB}{0, 0, 255}
\definecolor{codegreen}{rgb}{0,0.6,0}
\definecolor{codegray}{rgb}{0.5,0.5,0.5}
\definecolor{codepurple}{rgb}{0.58,0,0.82}
\definecolor{hotpink}{RGB}{255, 105, 180}

\newcommand{\vcenteredinclude}[1]{\begingroup
\setbox0=\hbox{\includegraphics[height=1.0em]{#1}}%
\parbox{\wd0}{\box0}\endgroup}
\newcommand{\xxmark}{\vcenteredinclude{figures/xmark.pdf}}
\newcommand{\ccmark}{\vcenteredinclude{figures/cmark.pdf}}
\newcommand{\ycmark}{\vcenteredinclude{figures/yellow_cmark.pdf}}
\newcommand{\yxmark}{\vcenteredinclude{figures/yellow_xmark.pdf}}
\newcommand\mybox[2][]{\tikz[overlay]\node[inner sep=1pt, anchor=text, rectangle, rounded corners=1mm,#1] {#2};\phantom{#2}}
\definecolor{fillcolor}{RGB}{216,217,252}
\newcommand\goodratspan[1]{\mybox[fill=cyan!15]{#1}}
\newcommand\badratspan[1]{\mybox[fill=red!15]{#1}}
\newcommand\tracespan[1]{\mybox[fill=yellow!20]{#1}}
\newcommand\trace[1]{{\tt \textcolor{magenta}{Trace #1}}}

%\titlespacing*{\section}{1pt}{0.5\baselineskip}{0.5\baselineskip}

\setminted{fontsize=\scriptsize}
\newenvironment{promptfence}{\captionsetup{type=listing}}{}

\newcommand{\xxcomment}[3]{\textcolor{#1}{[#2 #3]}}

\newcommand{\robin}[1]{
\xxcomment{purple}{Robin:}{ #1}
}

\newcommand{\jinjun}[1]{
 \xxcomment{blue}{Jinjun:}{ #1}
}

\newcommand{\gail}[1]{
 \xxcomment{green}{Gail:}{ #1}
}

\newcommand{\junfeng}[1]{
 \xxcomment{green}{Junfeng:}{ #1}
}

\newcommand{\bray}[1]{
 \xxcomment{red}{Baishakhi:}{ #1}
}

\newcommand{\rayb}[1]{
 \xxcomment{red}{Baishakhi:}{ #1}
}

\newcommand{\ira}[1]{
 \xxcomment{hotpink}{Ira:}{ #1}
}

% \input{sections/1_introduction}
% --------- Main sections -----------
\section{Introduction}
\label{sec:intro}

Automated program repair (APR) agents are autonomous systems that leverage the reasoning and generative capabilities of Large Language Models (LLMs) to perform complex, multi-step software development tasks. To support such capabilities, APR agents incorporate planning, self-reflection, and external tool use, including web search, file I/O, static analysis, and debugging—enabling them to decompose high-level objectives, assess their own intermediate outputs, and interact with the broader development ecosystem. Recent research has shown APR agents to be successful in fixing real-world bugs~\cite{zhang2024autocoderover,xia2024agentless, wang2024openhands, pabba2025refineenhancingprogramrepair}, often utilizing sub-agents for tasks like code synthesis from natural language specifications~\cite{wang2024openhands}, fault localization~\cite{li2024llm}, and automated test generation~\cite{ahmed2025otter}.
 
Traditional repair systems typically rely on manually crafted templates~\cite{liu2019tbar}, predefined transformation rules, or search-guided enumeration of patches~\cite{le2019automated}. In contrast, APR agents offer a more general and adaptive approach. They can ingest structured or unstructured bug reports, localize faults using lightweight static analysis or test-driven diagnosis, synthesize candidate patches, validate them through iterative compilation and test execution, and ultimately decide when and how to apply a fix. 
This end-to-end workflow—spanning fault localization, patch synthesis, patch validation, and code integration—broadly mirrors cognitive processes of human developers, highlighting the potential of APR agents to be autonomous repair assistants. 
A leading benchmark for evaluating automated program repair agents is SWE-bench~\cite{jimenez2023swe}, comprising 2,294 real-world tasks curated from 12 actively maintained Python GitHub projects. Each task is anchored to a specific repository version and accompanied by a natural language issue description, requiring generation of a patch that resolves the issue. Evaluation is conducted via test execution, including fail-to-pass tests that confirm issue resolution and pass-to-pass tests that check for regression safety. To improve the robustness of the benchmark, OpenAI curated a high-quality subset called SWE-bench Verified~\cite{swebench-verified}, consisting of 500 tasks with well-defined and semantically meaningful test cases. 

SWE-bench and its Verified variant have quickly become the de facto standard for assessing LLM-based repair systems. They are widely used in academic research and industrial settings to evaluate models' code understanding, reasoning, and patch synthesis abilities. As a result, an increasing number of agents—developed across both academic and industrial settings—are being explicitly designed to target this benchmark, with growing emphasis on achieving steady improvements in the number of successfully repaired issues, as reflected in the evolving SWE-bench leaderboard~\cite{swebench_leaderboard}.

While agents have achieved impressive leaderboard scores, such metrics reveal little about \emph{how} agents reason, \emph{why} they fail, or \emph{how} their behavior compares to that of human developers. As agent development accelerates, there is growing need for analyses beyond pass/fail outcomes. This work addresses that gap by tracing agents’ complete decision-making pipelines—from issue description to patch validation—and structuring our investigation through the following research questions.

We begin by examining differences between agent-generated and developer-written patches to understand how agent solutions diverge from those of developers, and why these differences arise.

\textbf{RQ1: How similar are the agent-generated and developer written patches?}
We find even when an agent-generated patch passes the given test suite, it often diverges from the developer patch—typically being more verbose and structurally complex. This motivates a deeper study of agent trajectories to explore how agents generate patches and where decision-making breaks down.

\textbf{RQ2: Where do APR agents struggle to generate correct patches?} 
We find that, even when using a strong LLM (e.g., Claude 3.5 Sonnet), agents struggle to fix logic-intensive issues. 
Moreover, agents tend to overfit to the given fail-to-pass test cases---more test cases provide greater validation constraints, making it harder for agents to generate spurious fixes. 
These findings highlight the significance of patch validation through testing and motivate us to study it in depth. 

\textbf{RQ3: Do APR agents adequately validate generated patches?} 
Agents frequently fail to reproduce reported issues through generated test cases (i.e., fail-to-pass cases) and often misidentify appropriate regression tests (i.e., pass-to-pass cases). Our analysis shows enhanced localization and scalable inference methods have the potential to substantially improve test generation effectiveness.

Despite a long line of research on test generation and regression test selection~\cite{harrold2001regression,shi2019reflection,wang2024systematic, huang2009orts}, agents struggle with both tasks in practice. This observation led us to question what kinds of tools or mechanisms these agents actually employ, motivating the following research question (RQ).

\textbf{RQ4: What tools do APR agents use?} 
Most agents rely on simple bash scripts with limited testing and debugging support, pointing to an opportunity to equip APR agents with richer tooling.

Together, these findings reveal both the capabilities and the limitations of current APR agents, offering concrete guidance for developing next-generation systems with greater reliability, transparency, and alignment with developer practices. Advancing APR requires a \emph{shift-left} approach---a software development paradigm that moves quality assurance and testing ``left'' (earlier) in the project timeline, emphasizing high-quality test generation and validation in the early stages---to guide agents toward producing genuine rather than superficial plausible fixes. Future systems should further incorporate (1) richer and more integrated tool ecosystems, (2) diversified agentic architectures that leverage complementary strengths across agents, and (3) benchmarks that prioritize semantic repair quality and test generation fidelity over surface-level success metrics.
Rather than proposing new tools or automation, we aim to provide actionable insights through systematic empirical analysis of APR agents. The main contributions of this paper are as follows:

\begin{itemize}[noitemsep,topsep=0pt,leftmargin=*]
    \item \textbf{In-depth analysis.} We present the first systematic study of the behavior of LLM-based, automatic program repair agents on the SWE-bench benchmark, tracing their entire decision-making pipelines from issue description to patch application.
    \item \textbf{Taxonomy.} We develop a taxonomy of agent workflows, identifying core behavioral modules (e.g., bug localization, reproduction test generation, patch synthesis) and characterizing dominant design strategies.
    \item \textbf{Insights.} We offer insights for agent designers to produce more reliable and effective patches.
    \item \textbf{Reproduction Package.} We provide a replication package \cite{replicationPackage} with our framework and results.
\end{itemize}

\section{Study Methodology}
\label{sec:methodology}

This section outlines our approach for collecting and analyzing agent execution traces (Section~2.1) and presents our taxonomy of agentic architectures (Section~2.2).

\phantomsection\label{sec:trace}\noindent\textbf{\textsc{2.1 Trace Collection}}\quad For each repair attempt, we collect an execution trace: the set of observable artifacts produced by an agent on a given task. An execution trace comprises of: (1) the agent's generated patch, (2) the resolution verdict indicating whether the patch passes both the fail-to-pass tests that validate the fix and the pass-to-pass regression tests that ensure no existing functionality is broken, and (3) the action trajectory, the ordered sequence of interactions or actions taken by the agent during repair, capturing tool invocations, reasoning output, confirmations of bug reproductions, verifications of fix, and regression test selections. Together, these traces provide a multi-dimensional view of agent behavior, enabling analysis at both the outcome level (e.g., resolution rates, overfitting) and the process level (e.g., tool usage patterns).

Each trajectory consists of discrete steps, such as issuing commands, navigating files, editing code, or validating outputs.
For each action in the sequence, we extract the associated role label (e.g., \texttt{user}, \texttt{assistant}, \texttt{system}) along with a high-level summary of the LLM-generated response, capturing the agent's intended operation and rationale at each step. To illustrate, we present a short segment of an AutoCodeRover (ACR) trajectory\footnote{From the ACR + Claude-3.5-Sonnet trace for SWE-bench Verified issue \texttt{astropy\_\_astropy-7166}.}:

\definecolor{darkblue}{RGB}{0,0,120}
\definecolor{periwinkle}{RGB}{100,100,200}
\begin{tcolorbox}[colback=gray!10, colframe=gray!50, boxrule=0.5pt, left=4pt, right=4pt, top=4pt, bottom=4pt]
\scriptsize\ttfamily
\textcolor{darkblue}{\textbf{[\{"role":}} \textcolor{periwinkle}{\textbf{"system"}}\textcolor{darkblue}{\textbf{, "content":}} "You are a software developer maintaining a large project. You are working on an issue submitted to your project. The issue contains a description marked between <issue> and </issue>. Your task is to invoke a few search API calls to gather sufficient code context for resolving the issue..."\textcolor{darkblue}{\textbf{\},}}\\[3pt]
\textcolor{darkblue}{\textbf{\{"role":}} \textcolor{periwinkle}{\textbf{"user"}}\textcolor{darkblue}{\textbf{, "content":}} "<issue>InheritDocstrings metaclass doesn't work for properties. Inside the InheritDocstrings metaclass it uses inspect.isfunction which returns False for properties.</issue>"\textcolor{darkblue}{\textbf{\},}}\\[3pt]
\textcolor{darkblue}{\textbf{\{"role":}} \textcolor{periwinkle}{\textbf{"user"}}\textcolor{darkblue}{\textbf{, "content":}} "Based on the files, classes, methods, and code statements from the issue related to the bug, you can use the following search APIs to get more context of the project: search\_class(class\_name: str), search\_method(method\_name: str), search\_code(code\_str: str)..."\textcolor{darkblue}{\textbf{\},}}\\[3pt]
\textcolor{darkblue}{\textbf{\{"role":}} \textcolor{periwinkle}{\textbf{"assistant"}}\textcolor{darkblue}{\textbf{, "content":}} "Let me help search for relevant code context for this issue... Let's make these API calls: 1. search\_class("InheritDocstrings") 2. search\_code("inspect.isfunction")..."\textcolor{darkblue}{\textbf{\}]}}
\end{tcolorbox}

In this portion of a trajectory, each \texttt{\{"role": ..., "content": ...\}} object represents a single ``step'' in the trajectory. The role labels distinguish the agent's autonomous decisions (\texttt{assistant}) from the agent's prompts and environmental feedback (\texttt{user}).
We use an LLM\footnote{Claude Sonnet 4, used both for step-level annotation and LLM-based evaluation.} to annotate each step with a high-level summary of the action performed. For the trajectory above, this yields:

\begin{tcolorbox}[colback=gray!10, colframe=gray!50, boxrule=0.5pt, left=4pt, right=4pt, top=4pt, bottom=4pt]
\scriptsize\ttfamily
\textcolor{darkblue}{\textbf{Step 1}} (\textcolor{periwinkle}{\textbf{system}}): "Define agent role as software developer"\\[3pt]
\textcolor{darkblue}{\textbf{Step 2}} (\textcolor{periwinkle}{\textbf{user}}): "Present bug report about InheritDocstrings"\\[3pt]
\textcolor{darkblue}{\textbf{Step 3}} (\textcolor{periwinkle}{\textbf{user}}): "List available search APIs"\\[3pt]
\textcolor{darkblue}{\textbf{Step 4}} (\textcolor{periwinkle}{\textbf{assistant}}): "Formulate initial search strategy and request API calls"
\end{tcolorbox}

\noindent These step-level annotations make lengthy trajectories easier to interpret and verify manually. Although different agents produce distinct trajectories, they often perform semantically similar actions (e.g., retrieving context in step 4). By identifying these common patterns across agents, we manually derived generalized workflow nodes such as ``Provide Context.'' To ensure correctness and interpretability, we cross-validate extracted trajectories by comparing them against the open-source implementations and workflow diagrams published in each agent's respective documentation or paper. This validation allows us to confirm the trajectories reflect true agent behavior and that the summarized intents align with the agent's underlying design. In parallel, we extract agent-generated \texttt{patch.diff} files from traces and retrieve corresponding developer patches from the SWE-bench Verified Parquet dataset. We normalize both sets of patches by removing diff-specific symbols (e.g., "\texttt{+}", "\texttt{--}") and formatting artifacts, producing clean versions suitable for side-by-side comparison in downstream clone analysis and structural evaluation.

\phantomsection\label{sec:taxonomy}\noindent\textbf{\textsc{2.2 Constructing a Taxonomy}}\quad To better characterize each agent's unique capabilities and workflows, we construct a taxonomy of decision-making pathways based on their execution traces. The annotations of fine-grained agent steps (Section~2.1) are manually reviewed and corrected, beginning with a shared subset for calibration; disagreements are adjudicated by a third reviewer to ensure consistency before continuing independently. Annotators consisted of co-authors and external student annotators. Annotation was performed independently, with the same samples assigned to multiple annotators to avoid any single reviewer dominating the process. After independent annotation, observations were consolidated and disagreements were adjudicated by a third reviewer (a co-author), ensuring no single perspective determined the final taxonomy. All taxonomy nodes were cross-validated against each agent's published source code and documentation to ground the taxonomy in objective evidence rather than annotator interpretation. This follows a similar annotation methodology to those presented in Ma et al.~\cite{ma2024solidity} and Yu et al.~\cite{yu2024bugs}. We do not apply automated clustering; instead, we manually identify recurring patterns across agents (Section~2.1). For example, one agent might use a browser-based search after analyzing the repository structure, while another uses an internal-API at the same point. Though superficially different, both represent retrieving external information, so they merge into a single node (e.g., ``special commands/APIs''). If a node is rare but critical, we preserve it as a standalone. We cross-check source code and paper details to ensure no decision paths are distorted. Once verified, nodes and edges are added to the final graph, where edge weights indicate how many agents traverse each path.

As shown in \Cref{fig:agent-taxonomy}, we synthesize the behavior of five representative APR agents: AutoCodeRover, Agentless, SWE-Agent, OpenHands-CodeAct, and MASAI (see~\Cref{tab:agent-summary})—into a unified flow graph.
In this taxonomy, \textbf{nodes} represent discrete agent actions (e.g., code retrieval, patch synthesis, test invocation), and \textbf{edges} represent transitions between decisions or phases.
While all agents follow a shared high-level structure of project understanding, context construction, patch generation, and testing, they diverge in notable ways. OpenHands uniquely performs browser-based lookups. AutoCodeRover and Agentless restructure the codebase and employ keyword search for localization, while SWE-Agent relies on information retrieval. MASAI and Agentless both rank patch candidates, though MASAI does so only after generating a test template, reflecting a more staged validation strategy. Agents also differ in when they generate test cases: MASAI does so early in the pipeline, while SWE-Agent and OpenHands defer test generation until after patch synthesis.

The taxonomy highlights two strategies: \emph{iterative refinement}, as in OpenHands, SWE-Agent, and AutoCodeRover, which generate and refine a single patch; and \emph{parallel sampling}, as in MASAI and Agentless, which produce multiple candidates—only MASAI refines across them. Refinement occurs at multiple points, including file/class selection, context generation, and test generation.

Graph analysis identifies three core components across all agents: (1) \textbf{Localization}, the most connected node type and essential for context setup; (2) \textbf{Patch Generation}, a central phase; and (3) \textbf{Reproduction Testing}, necessary for validating correctness. These three components form the backbone of modern SWE-agent workflows and motivate deeper study, as presented in the following sections.
What we refer to as graph analysis involves identifying the frequently visited nodes across the taxonomy. We borrow the term ``frequently visited nodes'' from graph traversal algorithms to describe nodes that are reached often, which corresponds directly to having many incoming edges (high in-degree). Although identifying frequently visited nodes can be automated, given the manageable size of the graph and to ensure accuracy, we elect to perform this step manually. We count in-degrees of several nodes and note, for example, Special Commands/APIs has an in-degree of 8, Provide Context an in-degree of 6, Select Important Classes/Funcs an in-degree of 6, Edit Snippet an in-degree of 4, and Filter Patches an in-degree of 5. These frequently visited nodes (high in-degree components) indicate core functions around which we define workflow categories, arriving at mappings such as ``Localization'', ``Patch Generation'', and ``Reproduction Testing''. These high frequency nodes are used as a proxy for the core functional roles and represent the components that play a disproportionately important role across agent workflows.
\definecolor{ProjYellow}{RGB}{255,215,0}
\definecolor{ContextBlue}{RGB}{135,206,250}
\definecolor{SearchPurple}{RGB}{230,230,250}
\definecolor{TestGreen}{RGB}{193,240,193}
\definecolor{InputOrange}{RGB}{255,165,0}
\definecolor{OutputPink}{RGB}{255,182,193}
\begin{figure*}[htbp]
\centering
  \noindent
  \includegraphics[width=0.9\textwidth, trim=0 0 0 0, clip]{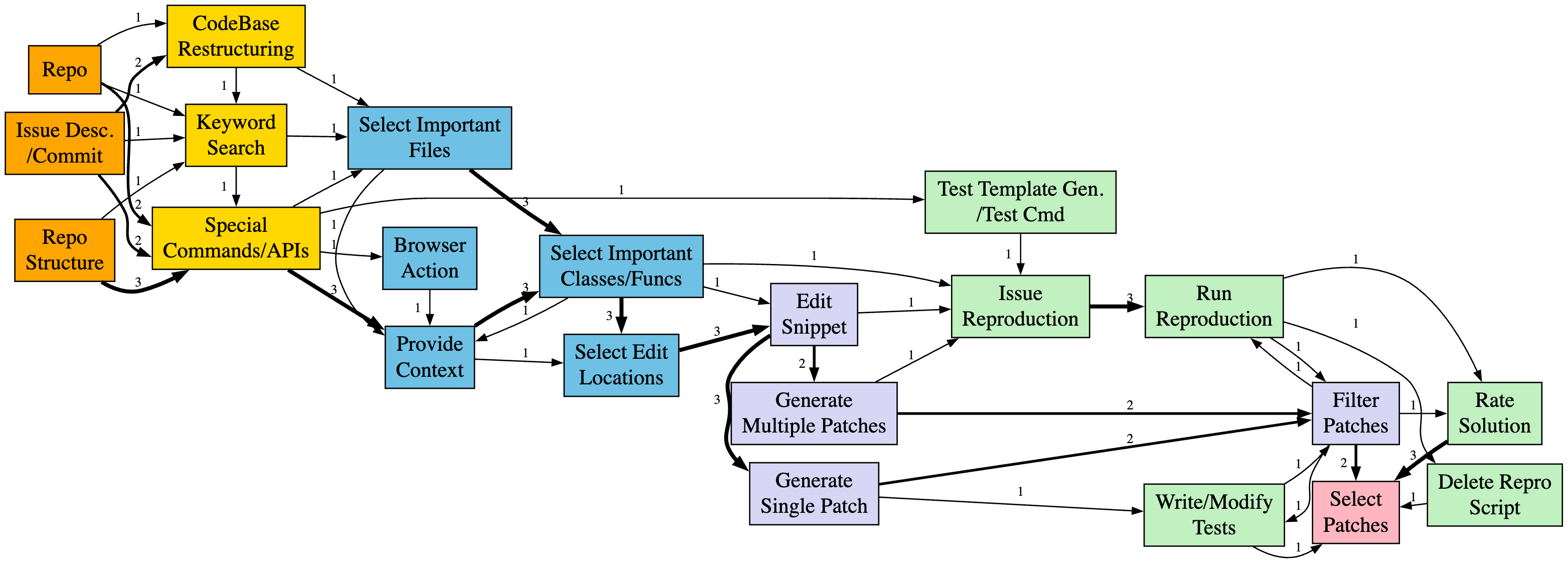}
  \caption{\textbf{Agent Taxonomy}. Nodes represent agent actions categorized by \colorbox{ProjYellow}{project understanding}, \colorbox{ContextBlue}{context understanding}, and \colorbox{SearchPurple}{patching} and \colorbox{TestGreen}{testing} actions. We highlight \colorbox{InputOrange}{input} and \colorbox{OutputPink}{output} nodes. Edges represent transitions, and edge weights represent the number of agents taking a particular path.}
  \label{fig:agent-taxonomy}
\end{figure*}

\section{Experimental Design}
\label{sec:experiment}

\phantomsection\label{sec:benchmark}\noindent\textbf{\textsc{3.1 Benchmark Selection}}\quad SWE-bench~\cite{jimenez2023swe} is a leading benchmark for evaluating APR agents, comprising 2,294 real-world tasks from 12 actively maintained Python GitHub repositories. Each task specifies a repository version and includes a natural language issue description, requiring a candidate patch that passes both \textit{fail-to-pass} (bug-fixing) and \textit{pass-to-pass} (regression) tests.

To improve benchmark reliability, OpenAI released \textit{SWE-bench Verified}~\cite{swebench-verified}, a curated subset of 500 tasks selected through a human annotation campaign involving 93 experienced developers. Annotators removed 68.3\% of tasks due to invalid test harnesses, vague descriptions, or environment dependencies. This subset serves as a high-quality gold standard for evaluating APR agents.

Each task in SWE-bench Verified is labeled with a difficulty level based on estimated human resolution time: \textbf{Easy} ($<$15 min), \textbf{Moderate} (15 min--1 hr), \textbf{Hard} (1--4 hr), and \textbf{Very Hard} ($>$4 hr). We use this difficulty signal to analyze agent performance and patch patterns. Agent logs, trajectories, and patch-level diffs are primarily retrieved from the official SWE-bench Experiments repository~\cite{swebenchExperimentsWebsite} using the AWS CLI. For configurations not available in that repository (e.g., OpenHands + GPT-4o), we use trajectories from ByteDance-Seed's Multi-SWE-bench\_trajs dataset~\cite{multiswebenchtrajs}. We further evaluate on \textit{Verified‑50} for controlled, computationally intensive experiments and manual analysis. This is an unbiased random subset of SWE‑bench Verified, introduced by Zainullina et al.~\cite{zainullina2025guided}. The subset offers a computationally efficient proxy for full benchmark evaluation and has been validated as representative of the full SWE-bench Verified benchmark, with results shown to match those on SWE‑bench Verified~\cite{verified50blog}.

\phantomsection\label{sec:agents}\noindent\textbf{\textsc{3.2 Agent Selection}}\quad We use the term ``APR agent'' to refer to any LLM-based system designed to solve automated program repair tasks—specifically, systems that generate code patches for natural language-described issues. Such agents augment a base LLM with tools for performing concrete actions such as navigating repositories, retrieving context, editing files, and running tests.

Broadly, we observe two kinds of agentic designs: (1) \textbf{Workflow-based agents} structure the repair process into a fixed pipeline of subtasks (e.g., localization, patch generation, validation), relying on pre-defined task flows and explicit module boundaries for modularity and reproducibility; (2) \textbf{Open-process agents} delegate more autonomy to the LLM by exposing it to a flexible interface (e.g., CLI, browser, file system) through which it can issue commands, make decisions interactively, and adaptively control execution, emphasizing flexibility and emergent strategies.

We select five agents representing these paradigms based on the following criteria: (1) strong performance on the SWE-bench leaderboard~\cite{swebench_leaderboard}; (2) open-source codebases or fully documented system descriptions; and (3) support for systematic inspection of decision trajectories.\footnote{We did not apply a fixed resolution-rate threshold; these agents were all among leading performers during the study period, with some reaching the top position at different points (e.g., OpenHands in April~\cite{openhands_sota_2025} and November 2025~\cite{swebench_leaderboard}, SWE-Agent in May 2025~\cite{swebench_leaderboard_2025}).} Importantly, the agents we study serve as foundations for many recent systems and benchmarks (e.g., SWE-bench Pro~\cite{deng2025swebenchpro} uses SWE-Agent scaffolding; SWE-RL~\cite{wei2025swerl} builds on Agentless), ensuring our findings remain relevant as the field evolves. \Cref{tab:agent-summary} summarizes the studied agents. We study agent trajectories generated with both Claude-3.5-Sonnet and GPT-4o. We selected agent-LLM configurations to maximize agent coverage using publicly available trajectories. Configurations are constrained by trajectory availability, as ACR+GPT-4o and MASAI+Claude-3.5-Sonnet were not publicly available. GPT-4o and Claude-3.5-Sonnet were two widely used, state-of-the-art backbones from two leading model families for which multiple agents had public trajectories, enabling controlled cross-agent comparisons. Our goal is to contrast the effects of different agent architectures and scaffolding designs rather than compare base LLM capabilities; thus, we hold the underlying model constant where possible. Since our findings target architectural behaviors rather than base model capability, they are more robust to rapid LLM progress.

We now present and analyze the results corresponding to our aforementioned research questions.

\begin{table}[t]
\centering
\caption{Summary of studied agents and their configurations.}
\label{tab:agent-summary}
\small
\begin{tabularx}{\textwidth}{@{} l X @{}}
\toprule
\textbf{Workflow-based Agent} & \textbf{Description} \\
\midrule
\textbf{Agentless (A)}~\cite{xia2024agentless} & Framework with a simple three-phase pipeline: localization, repair, and patch validation. We study Agentless + \textit{Claude-3.5-Sonnet} and \textit{GPT-4o}. \\
\addlinespace
\textbf{AutoCodeRover (ACR)}~\cite{zhang2024autocoderover} & Orchestrates LLM-driven subagents (AST retriever, fault localizer, patch generator) to fix GitHub issues. We study ACR + \textit{Claude-3.5-Sonnet}. \\
\addlinespace
\textbf{MASAI (M)}~\cite{arora2024masai} & Modular architecture with specialized sub-agents for information gathering and patch generation. We study MASAI + \textit{GPT-4o}. \\
\midrule
\textbf{Open-process Agent} & \textbf{Description} \\
\midrule
\textbf{SWE-Agent (SA)}~\cite{yang2024sweagent} & Equips LLM-based agents with an agent–computer interface for autonomous repository navigation, code editing, and test execution. We study SWE-Agent + \textit{Claude-3.5-Sonnet} and \textit{GPT-4o}. \\
\addlinespace
\textbf{OpenHands (OH)}~\cite{wang2024openhands} & Platform for flexible AI agents that write code, interact via command line and browser in sandboxed environments, and collaborate. We study OH + \textit{Claude-3.5-Sonnet} and \textit{GPT-4o}. \\
\bottomrule
\end{tabularx}
\end{table}

\section{{Agent vs.~Developer Written Patches (RQ1)}}
\label{sec:rq3}
\noindent\textbf{\textsc{4.1 Motivation}}\quad To understand how closely agent approaches align with human reasoning, we compare agent-generated patches to developer-written fixes. This moves beyond test-passing correctness to examine the semantic fidelity of agent solutions, whether they reflect the intent and minimalism often seen in human patches or rely on alternative, potentially overengineered strategies.\footnote{Consider a motivating example, \textit{astropy\_\_astropy-14096}: the developer applies a 7-line fix, reverting to default \texttt{\_\_getattribute\_\_} behavior, while an agent introduces layered fallback logic and explicit attribute checks totaling 53 lines. See \href{https://github.com/ARiSE-Lab/understanding-apr-agents/blob/main/submission/results/RQ1_motivating_ex.pdf}{motivating example}.} This divergence motivates a broader investigation: what patterns do agents introduce when solving real-world bugs, and how do they compare to human strategies in terms of complexity, generality, and semantic intent?

\noindent\textbf{\textsc{4.2 Approach}}\quad We evaluate similarities between developer and agent-generated patches using multiple complementary methods, analyzing characteristics such as diff size in lines, files edited, files added, reproducer files added, number of hunks, and average hunk size. For all experiments, we use agent trajectories generated with two backbones: Claude-3.5-Sonnet and GPT-4o. Edit-distance and clone analysis require resolved (correct) patches to enable meaningful comparison against developer solutions. For Claude-3.5-Sonnet, we analyze 911 of 918 resolved instances (99.2\%), excluding 7 OpenHands outliers exceeding the 200k-token limit. For GPT-4o, we analyze 587 of 588 resolved instances (99.8\%), excluding 1 outlier. We exclude very hard issues due to small sample size (3 issues) and near 0\% solve rates across nearly all agents and backbones. Edit patterns and hunk analysis examine patch characteristics regardless of correctness, reporting statistics on the maximal common subset of tasks where all agents using a given backbone generate patches. Manual analysis uses the Verified-50 subset and includes both correct and incorrect patches.

\textbf{A. Manual Patch Analysis.} We manually review patches for tasks in the Verified-50 dataset~\cite{zainullina2025guided}, a representative sample of 50 tasks from SWE-bench Verified. We analyze patches generated by one workflow-based agent (Agentless) and one open-process agent (SWE-Agent), comparing these to the developer-written solutions and to each other.

\textbf{B.~Code Clones. } We assess the similarity between agent-generated and developer-written patches using \textit{code clone} analysis, based on the taxonomy of Bellon et al.~\cite{bellon2007comparison}. Clones are categorized as follows: \textit{Type 1}—exact matches differing only in whitespace, layout, or comments; \textit{Type 2}—syntactically identical code with variations in identifiers, literals, or types; and \textit{Type 3}—structurally similar code with added, removed, or modified statements. As all studied patches pass at least one test case, they share a degree of semantic intent and can be treated as \textit{Type 4} clones by default. To classify clone types, we use Claude Sonnet 4 as an LLM-based judge, prompting it with a standardized template to evaluate each agent–developer patch pair.
A manual audit of 50 randomly selected patch pairs by two independent annotators shows strong agreement with human judgment (Cohen's \(\kappa = 0.81\)),\footnote{\href{https://github.com/ARiSE-Lab/understanding-apr-agents/blob/main/submission/results/T8_cohens_k.md}{Cohen's kappa validation results.}} supporting the reliability of this approach.

\textbf{C. Patch Edit-Distance.} To triangulate the LLM-based clone analysis, we compute the character-level Levenshtein distance \cite{Wagner1974EditDistance} between each successful agent-generated patch and its corresponding developer patch, grouping results by agent. The Levenshtein distance measures the minimum number of single-character edits (insertions, deletions, or substitutions) required to transform one string into another, providing a quantitative measure of textual similarity between patches.

\textbf{D. Edit Patterns and Hunk Analysis.} We analyze and compare several patch characteristics for over 400 SWE-bench Verified tasks. We examine ground-truth developer solutions alongside patches from workflow-based agents (AutoCodeRover, Agentless, MASAI) and open-process agents (SWE-Agent, OpenHands). For each backbone, we compute per-agent averages for the statistics in Table~\ref{tab:patch-stats} (diff size in lines, files edited, files added, reproducer files added, number of hunks, and average hunk size), on the maximal common subset of tasks where patches are generated for all agents using that backbone. A \textit{hunk} is a contiguous group of differing lines identified when comparing two files; the diff utility finds sequences of common lines interspersed with these hunks of differences.\footnote{\href{https://www.gnu.org/software/diffutils/manual/html_node/Hunks.html}{GNU Diffutils Manual: Hunks.}} We use the Python library \texttt{unidiff.PatchSet} to extract hunks from patches.

\begin{figure*}[!t]
  \centering

  \begin{minipage}[t]{0.30\textwidth}
    \centering
    \begin{tikzpicture}[baseline=(current bounding box.north)]
      \definecolor{gpt4ocolor}{RGB}{65,105,225}
      \definecolor{claudecolor}{RGB}{220,20,60}

      \begin{axis}[
          boxplot/draw direction=y,
          ylabel={Edit Distance (chars)},
          ylabel style={font=\small},
          xtick={1,2,3,4,5},
          xticklabels={ACR,A,OH,SA,M},
          x tick label style={font=\footnotesize},
          ymajorgrids=true,
          grid style=dashed,
          height=4.5cm,
          width=\linewidth,
          xmin=0.5,
          xmax=5.5,
          ymin=0,
          ymax=5000,
          title={(a) Patch Edit Distance},
          title style={font=\bfseries\small, yshift=2pt},
          boxplot/box extend=0.22,
      ]

      \addplot+[
        fill=claudecolor!25,
        draw=claudecolor,
        line width=1pt,
        forget plot,
        boxplot prepared={draw position=1,lower whisker=32,lower quartile=140,median=286,upper quartile=618,upper whisker=1330},
      ] coordinates {};

      \addplot+[
        fill=gpt4ocolor!25,
        draw=gpt4ocolor,
        line width=1pt,
        forget plot,
        boxplot prepared={draw position=1.8,lower whisker=30,lower quartile=96,median=269,upper quartile=612,upper whisker=1362},
      ] coordinates {};
      \addplot+[
        fill=claudecolor!25,
        draw=claudecolor,
        line width=1pt,
        forget plot,
        boxplot prepared={draw position=2.2,lower whisker=30,lower quartile=127,median=345,upper quartile=669,upper whisker=1480},
      ] coordinates {};

      \addplot+[
        fill=gpt4ocolor!25,
        draw=gpt4ocolor,
        line width=1pt,
        forget plot,
        boxplot prepared={draw position=2.8,lower whisker=36,lower quartile=1081,median=1582,upper quartile=2439,upper whisker=4475},
      ] coordinates {};
      \addplot+[
        fill=claudecolor!25,
        draw=claudecolor,
        line width=1pt,
        forget plot,
        boxplot prepared={draw position=3.2,lower whisker=36,lower quartile=510,median=1528,upper quartile=3207,upper whisker=4925},
      ] coordinates {};

      \addplot+[
        fill=gpt4ocolor!25,
        draw=gpt4ocolor,
        line width=1pt,
        boxplot prepared={draw position=3.8,lower whisker=33,lower quartile=163,median=402,upper quartile=2038,upper whisker=4847},
      ] coordinates {};
      \addplot+[
        fill=claudecolor!25,
        draw=claudecolor,
        line width=1pt,
        boxplot prepared={draw position=4.2,lower whisker=37,lower quartile=860,median=1467,upper quartile=2497,upper whisker=4930},
      ] coordinates {};

      \addplot+[
        fill=gpt4ocolor!25,
        draw=gpt4ocolor,
        line width=1pt,
        boxplot prepared={draw position=5,lower whisker=120,lower quartile=288,median=530,upper quartile=788,upper whisker=1445},
      ] coordinates {};

      \end{axis}

      \node[anchor=north] at (1.2,-0.2) {
        \begin{tikzpicture}[scale=0.7]
          \draw[gpt4ocolor, line width=0.8pt] (0,0) -- (0,0.4);
          \draw[gpt4ocolor, fill=gpt4ocolor!25, line width=0.8pt] (-0.08,0.1) rectangle (0.08,0.3);
          \draw[gpt4ocolor, line width=0.8pt] (-0.08,0.2) -- (0.08,0.2);
          \node[anchor=west, font=\scriptsize] at (0.15,0.2) {G};

          \draw[claudecolor, line width=0.8pt] (0.8,0) -- (0.8,0.4);
          \draw[claudecolor, fill=claudecolor!25, line width=0.8pt] (0.72,0.1) rectangle (0.88,0.3);
          \draw[claudecolor, line width=0.8pt] (0.72,0.2) -- (0.88,0.2);
          \node[anchor=west, font=\scriptsize] at (0.95,0.2) {C};
        \end{tikzpicture}
      };
    \end{tikzpicture}
  \end{minipage}
  \hfill
  \begin{minipage}[t]{0.34\textwidth}
    \centering
    \vspace{0.1cm}
    {\bfseries\small (b) Clone Distribution (\%)}
    \vspace{0.3em}

    \footnotesize
    \setlength{\tabcolsep}{3pt}
    \begin{tabular}{ll|rrrr}
      \toprule
      \textbf{Agent} & & \textbf{T1} & \textbf{T2} & \textbf{T3} & \textbf{N/A} \\
      \midrule
      ACR & C & 26.0 & 12.6 & 18.2 & 43.3 \\
      \midrule
      \rowcolor{gray!10} A & C & 26.8 & 8.7 & 18.1 & 46.5 \\
      \rowcolor{gray!10}   & G & 27.1 & 9.9 & 14.4 & 48.6 \\
      \midrule
      OH & C & 14.3 & 8.9 & 20.5 & 56.2 \\
         & G & 7.1 & 8.7 & 16.5 & 67.7 \\
      \midrule
      \rowcolor{gray!10} SA & C & 6.5 & 9.5 & 17.3 & 66.7 \\
      \rowcolor{gray!10}    & G & 19.8 & 10.3 & 16.4 & 53.4 \\
      \midrule
      M & G & 17.2 & 13.5 & 20.2 & 49.1 \\
      \bottomrule
    \end{tabular}

    \vspace{0.2em}
    \tiny C = Claude-3.5, G = GPT-4o
  \end{minipage}
  \hfill
  \begin{minipage}[t]{0.34\textwidth}
    \centering
    \vspace{0.1cm}
    {\bfseries\small (c) Clone Breakdown (Easy)}

    \vspace{-0.1cm}
    \begin{tikzpicture}[baseline=(current bounding box.north)]
      \definecolor{cgray}{RGB}{100,100,100}   % n/a
      \definecolor{cblue}{RGB}{170,70,70}     % Type-1
      \definecolor{cyellow}{RGB}{205,178,102} % Type-2
      \definecolor{corange}{RGB}{217,131,74}  % Type-3

      \begin{axis}[
          axis line style=draw=none,
          xbar stacked,
          y=0.42cm,
          width=\linewidth,
          height=4.5cm,
          xmin=0, xmax=100,
          scaled x ticks=false,
          xtick={0,50,100},
          xticklabels={0\%,50\%,100\%},
          xticklabel style={font=\scriptsize},
          symbolic y coords={M-G, SA-G, SA-C, OH-G, OH-C, ACR-C, A-G, A-C},
          ytick=data,
          yticklabel style={font=\scriptsize},
          y dir=reverse,
          enlarge y limits=0.1,
          enlarge x limits={abs=3},
          nodes near coords,
          nodes near coords align={center},
          every node near coord/.append style={font=\tiny},
          legend columns=4,
          legend cell align=left,
          legend style={
            at={(0.5,-0.12)},
            anchor=north,
            font=\scriptsize,
            column sep=1pt,
            draw=none,
            fill=none,
          },
          legend image code/.code={
            \draw[draw=none,fill=#1] (0cm,-0.08cm) rectangle (0.2cm,0.08cm);
          }
      ]
          \addplot+[xbar, fill=cgray!90!white, draw=none, every node near coord/.append style={text=white}] coordinates { (49.5,M-G) (53.5,SA-G) (68.8,SA-C) (68.0,OH-G) (51.1,OH-C) (35.9,ACR-C) (41.1,A-G) (32.8,A-C) };
          \addplot+[xbar, fill=cblue!90!white, draw=none, every node near coord/.append style={text=white}] coordinates { (23.7,M-G) (26.8,SA-G) (8.6,SA-C) (9.3,OH-G) (21.5,OH-C) (35.2,ACR-C) (38.3,A-G) (39.4,A-C) };
          \addplot+[xbar, fill=cyellow!80!white, draw=none, every node near coord/.append style={text=black}] coordinates { (9.7,M-G) (8.5,SA-G) (8.6,SA-C) (8.0,OH-G) (8.1,OH-C) (13.3,ACR-C) (9.3,A-G) (11.7,A-C) };
          \addplot+[xbar, fill=corange!90!white, draw=none, every node near coord/.append style={text=black}] coordinates { (17.2,M-G) (11.3,SA-G) (14.0,SA-C) (14.7,OH-G) (19.3,OH-C) (15.6,ACR-C) (11.2,A-G) (16.1,A-C) };
          \legend{N/A, T1, T2, T3}
      \end{axis}
    \end{tikzpicture}
  \end{minipage}

  \caption{
    \textbf{Patch similarity across agents.}
    (a) Edit distance between developer and agent patches.
    (b) Clone distribution for resolved patches.
    (c) Clone breakdown for easy tasks.\protect\footnotemark{}
    C = Claude-3.5, G = GPT-4o; T1/T2/T3 = Type-1/2/3 clones; N/A = not a clone. ACR uses C only; MASAI uses G only.
  }
  \label{fig:rq3-threepanel}
\end{figure*}
\footnotetext{\protect\href{https://github.com/ARiSE-Lab/understanding-apr-agents/blob/main/submission/results/T4_clone_detection_similarity.md}{Full clone detection and edit distance results, all difficulty levels.}}

\noindent\textbf{\textsc{4.3 Results}}\quad \textbf{Manual patch analysis trends.} When both SWE-Agent and Agentless successfully resolve an issue, Agentless often produces a shorter solution. However, Agentless struggles with multi-file edits or complex code changes like updating data structures, sometimes generating correct patches in one location while missing other bug locations. Conversely, SWE-Agent's tendency toward larger structural changes enables it to resolve tasks requiring extensive edits (e.g., adding new methods), but its patches are often unnecessarily complex, longer than developer solutions, and span more files. For tasks with simple fixes, SWE-Agent's verbose patches are harder for developers to interpret, evaluate, and maintain if merged.
These observations confirm quantitative findings: SWE-Agent generates longer, multi-file edits (partly due to test additions and added complexity), while Agentless generates shorter, localized single-file edits more similar to developer solutions.\footnote{\href{https://github.com/ARiSE-Lab/understanding-apr-agents/blob/main/submission/results/nebius-50-annotations.pdf}{Full Verified-50 annotations.}}

\textbf{Patch similarity based on edit distance.}  Fig. \ref{fig:rq3-threepanel}(a) presents the character-level edit distance between agent-generated and developer-written patches across five agents. The boxplots reveal distinct patterns across agents with consistent trends across backbones. Workflow-based agents exhibit lower edit distances: AutoCodeRover (Claude-3.5) has a median of 286 characters, Agentless 269 (GPT-4o) and 345 (Claude-3.5), and MASAI (GPT-4o) 530. Open-process agents exhibit higher edit distances: OpenHands 1,582 (GPT-4o) and 1,528 (Claude-3.5), SWE-Agent 1,467 (Claude-3.5) and 402 (GPT-4o). Although SWE-Agent's GPT-4o median is lower, its distribution is notably wider, and at least half of patches still exceed 402 characters, indicating greater variability in how closely patches match developer solutions. However, edit distance is inherently noisy, as it measures textual similarity rather than functional equivalence. Large character differences may reflect alternative but equally valid implementations. To address these limitations, we complement this quantitative analysis with a qualitative code clone study (Fig. \ref{fig:rq3-threepanel}(b-c)) that examines structural and semantic similarities beyond surface-level text matching. These patterns align with the clone analysis that follows: lower edit distance tends to coincide with higher Type-1/2 clone rates, while larger edit distance tends to coincide with higher non-clone rates.

\definecolor{cgray}{RGB}{100,100,100}
\definecolor{cblue}{RGB}{86,119,155}
\definecolor{corange}{RGB}{217,131,74}
\definecolor{cyellow}{RGB}{205,178,102}

\textbf{Patch similarity based on code clones.} Fig. \ref{fig:rq3-threepanel}(b) shows clone-type distributions. Overall, non-clones dominate across all agents (43--68\%), indicating that many successful patches follow approaches structurally distinct from the developer fix. Among clone cases, Type-1 and Type-3 are both common (Type-1: 6.5--27.1\%; Type-3: 14.4--20.5\%), while Type-2 is comparatively uncommon (8.7--13.5\%). Agent variation is substantial: ACR and Agentless exhibit higher Type-1 rates (26.0--27.1\%), while OpenHands produces the highest non-clone rates (56.2--67.7\%), suggesting more divergent repair strategies on average. Fig. \ref{fig:rq3-threepanel}(c) breaks down clone types by issue difficulty. Workflow-based agents such as ACR, Agentless, and MASAI show a clear tendency to produce \textit{Type-1} and \textit{Type-2} clones on easier issues, reflecting a higher degree of syntactic overlap with developer patches. 
For instance, on easy tasks, the combined share of \textit{Type-1/2} clones reaches 48.5\% for ACR (Claude-3.5), 51.1\% for Agentless (Claude-3.5), and 33.4\% for MASAI (GPT-4o), compared to 17.2\% for SWE-Agent (Claude-3.5) and 29.6\% for OpenHands (Claude-3.5). 
This pattern suggests workflow-driven repair pipelines are more likely to replicate developer-like edits when problems are straightforward and well-structured. 
As task difficulty increases, clone similarity drops across all agents and the proportion of “n/a” (non-clone) instances rises sharply, indicating harder bugs require more structural rethinking and lead to functionally correct but syntactically distinct repairs. 

\textbf{Edit patterns and hunk analysis.} Table~\ref{tab:patch-stats} shows a clear distinction of agent styles across both backbones: workflow agents produce concise, localized edits, while open-process agents make broader changes spanning more files and edit locations. Agentless exemplifies the workflow pattern, averaging 22--23 changed lines (Claude-3.5: 22.0; GPT-4o: 22.6), editing roughly one file (1.01; 1.09) with 1.3 hunks (1.32; 1.37) of about 12 lines each (12.14; 11.82). Open-process agents diverge substantially: SWE-Agent averages 101.5--245.7 lines changed, 3.5+ hunks of 21--24 lines, and adds 1.6--2.0 files per patch. OpenHands produces the largest diffs (474.2; 331.2 lines), edits the most files (5.18; 1.57), and adds multiple files (3.97; 1.92). OpenHands adds only 0.06--1.08 reproducer files on average, indicating most added files are non-test code. Developer patches are similarly concise for easy and moderate tasks ($\approx$ 20–40 lines) but scale more predictably with difficulty ($\approx$ 6× increase from easy to hard). Workflow agents mirror this scaling modestly. Developers make smaller edits across more locations, averaging 2–3 hunks across 1–2 files, situating their patch profiles between the focused edits of workflow agents and the expansive modifications of open-process agents.

\begin{findingbox}

\textbf{RQ1 Findings:} Workflow-based agents produce concise, developer-aligned patches (35--41\% Type-1/2 clones; single-file edits averaging 22--23 lines), while open-process agents generate solutions up to 40$\times$ larger that span multiple files. Clone similarity decreases with task difficulty, with 43--68\% of successful patches being non-clones; these patterns hold across backbones, corroborated by edit distance and manual analysis.

\end{findingbox}

\begin{table*}[!t]
\centering
\caption{\textbf{Patch Analysis.} For each agent and developer patches, we report average: edited lines, edited files, added files, added reproducer files, hunks, and hunk size (lines). C = Claude-3.5, G = GPT-4o. ACR uses C only; MASAI uses G only. — = no data for that backbone. Gold/Developer = ground-truth human patches as baseline; C/G differ due to maximal common task subsets per backbone (424/415).\protect\footnotemark}
\scriptsize
\setlength{\tabcolsep}{2.5pt}
\begin{tabular}{
  l|l|rr|rr|rr|rr|rr|rr|rr
}
\toprule
\textbf{Agent}
  & \textbf{Difficulty}
  & \multicolumn{2}{c|}{\textbf{\# Tasks}}
  & \multicolumn{2}{c|}{\makecell{\textbf{Diff Size}\\\textbf{(\# Lines)}}}
  & \multicolumn{2}{c|}{\makecell{\textbf{\# Files}\\\textbf{Edited}}}
  & \multicolumn{2}{c|}{\makecell{\textbf{\# Files}\\\textbf{Added}}}
  & \multicolumn{2}{c|}{\makecell{\textbf{\# Reproducer}\\\textbf{Files Added}}}
  & \multicolumn{2}{c|}{\textbf{\# Hunks}}
  & \multicolumn{2}{c}{\textbf{Hunk Size}}
  \\
  &
  & \textbf{C} & \textbf{G}
  & \textbf{C} & \textbf{G}
  & \textbf{C} & \textbf{G}
  & \textbf{C} & \textbf{G}
  & \textbf{C} & \textbf{G}
  & \textbf{C} & \textbf{G}
  & \textbf{C} & \textbf{G}
  \\
\hline
% Gold / Developer
\rowcolor{gray!10}
\cellcolor{white}
  & All       & 424 & 415 & 38.7  & 38.0  & 1.26 & 1.25 & 0    & 0    & 0    & 0    & 2.51 & 2.40 & 12.54 & 12.93 \\
  & Easy      & 165 & 162 & 19.5  & 18.8  & 1.03 & 1.02 & 0    & 0    & 0    & 0    & 1.39 & 1.35 & 10.15 & 10.05 \\
  \rowcolor{gray!10}
  \cellcolor{white}
  & Moderate  & 223 & 214 & 38.5  & 37.9  & 1.30 & 1.28 & 0    & 0    & 0    & 0    & 2.52 & 2.43 & 13.64 & 14.27 \\
\multirow{-4}{*}{Gold / Developer}
  & Hard      & 34  & 36  & 115.9 & 102.6 & 2.11 & 2.03 & 0.03 & 0.03 & 0    & 0    & 6.41 & 5.31 & 16.73 & 17.69 \\
\hline
% AutoCodeRover (Claude-3.5 only)
  & All       & 424 & --  & 23.0  & --    & 1.03 & --   & 0    & --   & 0    & --   & 1.25 & --   & 13.47 & -- \\
  \rowcolor{gray!10}
  \cellcolor{white}
  & Easy      & 165 & --  & 19.5  & --    & 1.02 & --   & 0    & --   & 0    & --   & 1.20 & --   & 11.07 & -- \\
  & Moderate  & 223 & --  & 23.8  & --    & 1.02 & --   & 0    & --   & 0    & --   & 1.25 & --   & 14.38 & -- \\
  \rowcolor{gray!10}
  \cellcolor{white}
\multirow{-4}{*}{AutoCodeRover}
  & Hard      & 34  & --  & 35.0  & --    & 1.11 & --   & 0    & --   & 0    & --   & 1.58 & --   & 19.38 & -- \\
\hline
% Agentless (both backbones)
\rowcolor{gray!10}
\cellcolor{white}
  & All       & 424 & 415 & 22.0  & 22.6  & 1.01 & 1.09 & 0    & 0    & 0    & 0    & 1.32 & 1.37 & 12.14 & 11.82 \\
  & Easy      & 165 & 162 & 19.4  & 19.6  & 1.00 & 1.02 & 0    & 0    & 0    & 0    & 1.27 & 1.27 & 10.49 & 10.50 \\
  \rowcolor{gray!10}
  \cellcolor{white}
  & Moderate  & 223 & 214 & 21.9  & 23.5  & 1.01 & 1.12 & 0    & 0    & 0    & 0    & 1.24 & 1.38 & 12.97 & 12.36 \\
\multirow{-4}{*}{Agentless}
  & Hard      & 34  & 36  & 33.2  & 28.9  & 1.08 & 1.28 & 0    & 0    & 0    & 0    & 1.76 & 1.61 & 14.89 & 14.06 \\
\hline
% SWE-Agent (both backbones)
  & All       & 424 & 415 & 101.5 & 245.7 & 2.96 & 1.22 & 1.60 & 1.97 & 0.15 & 0.62 & 3.62 & 3.52 & 23.83 & 21.30 \\
  \rowcolor{gray!10}
  \cellcolor{white}
  & Easy      & 165 & 162 & 69.6  & 353.7 & 2.55 & 1.26 & 1.15 & 1.82 & 0.10 & 0.54 & 3.00 & 3.32 & 17.34 & 22.86 \\
  & Moderate  & 223 & 214 & 119.3 & 162.8 & 3.15 & 1.18 & 1.82 & 1.76 & 0.19 & 0.64 & 3.80 & 3.24 & 28.10 & 19.74 \\
  \rowcolor{gray!10}
  \cellcolor{white}
\multirow{-4}{*}{SWE-Agent}
  & Hard      & 34  & 36  & 133.2 & 264.4 & 3.35 & 1.25 & 2.02 & 4.03 & 0.11 & 0.89 & 5.00 & 5.75 & 28.00 & 24.43 \\
\hline
% OpenHands (both backbones)
\rowcolor{gray!10}
\cellcolor{white}
  & All       & 424 & 415 & 474.2 & 331.2 & 5.18 & 1.57 & 3.97 & 1.92 & 0.06 & 1.08 & 4.97 & 3.15 & 30.53 & 21.14 \\
  & Easy      & 165 & 162 & 857.2 & 405.8 & 8.58 & 2.28 & 7.49 & 1.78 & 0.09 & 1.04 & 6.84 & 2.98 & 24.38 & 19.20 \\
  \rowcolor{gray!10}
  \cellcolor{white}
  & Moderate  & 223 & 214 & 209.8 & 299.3 & 3.00 & 1.09 & 1.76 & 2.03 & 0.05 & 1.01 & 3.48 & 3.25 & 32.07 & 22.77 \\
\multirow{-4}{*}{OpenHands}
  & Hard      & 34  & 36  & 259.7 & 204.6 & 3.08 & 1.33 & 1.47 & 2.00 & 0    & 1.69 & 5.41 & 3.53 & 44.00 & 20.80 \\
\hline
% MASAI (GPT-4o only)
  & All       & --  & 415 & --    & 29.1  & --   & 1.05 & --   & 0    & --   & 0    & --   & 1.89 & --    & 13.57 \\
  \rowcolor{gray!10}
  \cellcolor{white}
  & Easy      & --  & 162 & --    & 23.1  & --   & 1.03 & --   & 0    & --   & 0    & --   & 1.56 & --    & 12.24 \\
  & Moderate  & --  & 214 & --    & 31.3  & --   & 1.06 & --   & 0    & --   & 0    & --   & 2.01 & --    & 14.19 \\
  \rowcolor{gray!10}
  \cellcolor{white}
\multirow{-4}{*}{MASAI}
  & Hard      & --  & 36  & --    & 41.2  & --   & 1.08 & --   & 0    & --   & 0    & --   & 2.42 & --    & 15.84 \\
\bottomrule
\end{tabular}
\label{tab:patch-stats}
\end{table*}
\footnotetext{\protect\href{https://github.com/ARiSE-Lab/understanding-apr-agents/blob/main/submission/results/rq1_patch_analysis_full_table.pdf}{Full table with Very Hard issues (max s=3).}}

\section{Agent Effectiveness in Patch Generation (RQ2)}
\noindent\textbf{\textsc{5.1 Motivation}}\quad We are motivated to examine APR agents' issue-fixing capabilities, identifying both where they excel and struggle. This analysis helps surface both the strengths and limitations of these systems and highlights opportunities for advancing their design and effectiveness.

\noindent\textbf{\textsc{5.2 Approach}}\quad Using agent trajectories from Claude-3.5-Sonnet and GPT-4o backbones, we evaluate performance across difficulty levels, analyze capability divergence, and investigate the impact of test coverage on patch correctness through three research questions:
\vspace{-0.1em}
\begin{itemize}[leftmargin=*]
    \item \textbf{RQ2.1~Performance across difficulty levels:} We examine how agent performance varies across issues of increasing complexity, using the difficulty tiers defined in SWE-bench Verified (Section~3.2). This assesses how well agents scale from simple, syntactic fixes to semantically complex and time-consuming challenges.

    \item \textbf{RQ2.2~Divergence in agent capability:} We analyze whether agents solve overlapping or disjoint issue subsets and identify which bug types each uniquely solves. We report the moderate tier as it offers the most informative comparison with sufficient sample size and mixed solve rates; easy and hard tiers show uniformly high and low rates, respectively, limiting discriminative insights.\footnote{\href{http://github.com/ARiSE-Lab/understanding-apr-agents/blob/main/submission/results/T3_agent_strengths.md}{Unique strengths across backbone comparisons and difficulty tiers.}} This reveals if agents exhibit complementary strengths useful for ensemble approaches.

    \item \textbf{RQ2.3~Overfitting and test case coverage:} We explore the effect of fail-to-pass test case availability on patch success rates and overfitting behavior. Specifically, we ask if having more fail-to-pass tests correlates with lower resolution due to more rigorous validation constraints. By manual analysis of agent trajectories, we quantify overfitting rates where agents produce patches that resolve agent-generated reproducers but fail actual validation. This evaluates robustness of agent capabilities and highlights the role of test coverage in enforcing semantic correctness.
\end{itemize}

\begin{figure*}[!t]
\centering
\hspace*{-1.8em}
\begin{minipage}[t]{0.30\textwidth}
\centering
\footnotesize
\setlength{\tabcolsep}{3pt}
\renewcommand{\arraystretch}{1.1}
{\bfseries\small (a) Resolution by Difficulty}
\vspace{0.3em}

\begin{tabular}{l|cccc}
\toprule
& \textbf{Easy} & \textbf{Mod.} & \textbf{Hard} & \textbf{V.H.} \\
\midrule
\rowcolor{gray!15}\multicolumn{5}{l}{\textit{GPT-4o}} \\
\textbf{M}   & \cellcolor{green!40}47.9 & \cellcolor{green!25}26.1 & \cellcolor{green!10}4.8 & \cellcolor{red!30}0 \\
\textbf{SA}  & \cellcolor{green!30}36.6 & \cellcolor{green!15}16.9 & \cellcolor{green!5}2.4 & \cellcolor{red!30}0 \\
\textbf{A}   & \cellcolor{green!50}55.2 & \cellcolor{green!25}28.0 & \cellcolor{green!5}2.4 & \cellcolor{red!30}0 \\
\textbf{OH}  & \cellcolor{green!30}38.7 & \cellcolor{green!15}19.5 & \cellcolor{green!10}4.8 & \cellcolor{red!30}0 \\
\midrule
\rowcolor{gray!15}\multicolumn{5}{l}{\textit{Claude-3.5-Sonnet}} \\
\textbf{ACR} & \cellcolor{green!75}66.0 & \cellcolor{green!30}38.3 & \cellcolor{green!15}7.1 & \cellcolor{red!30}0 \\
\textbf{SA}  & \cellcolor{green!40}47.9 & \cellcolor{green!25}28.0 & \cellcolor{green!10}4.8 & \cellcolor{red!30}0 \\
\textbf{A}   & \cellcolor{green!80}70.6 & \cellcolor{green!35}42.5 & \cellcolor{green!20}14.3 & \cellcolor{red!30}0 \\
\textbf{OH}  & \cellcolor{green!100}72.7 & \cellcolor{green!40}44.1 & \cellcolor{green!25}19.0 & \cellcolor{green!30}33.3 \\
\midrule
\textbf{N} & \textbf{194} & \textbf{261} & \textbf{42} & \textbf{3} \\
\bottomrule
\end{tabular}
\label{tab:issue-counts}
\end{minipage}
\hspace{-0.85em}
\begin{minipage}[t]{0.31\textwidth}
\centering
{\bfseries\small (b) Agent Overlap (Moderate)}
\vspace{0.3em}

\includegraphics[width=1.1\linewidth]{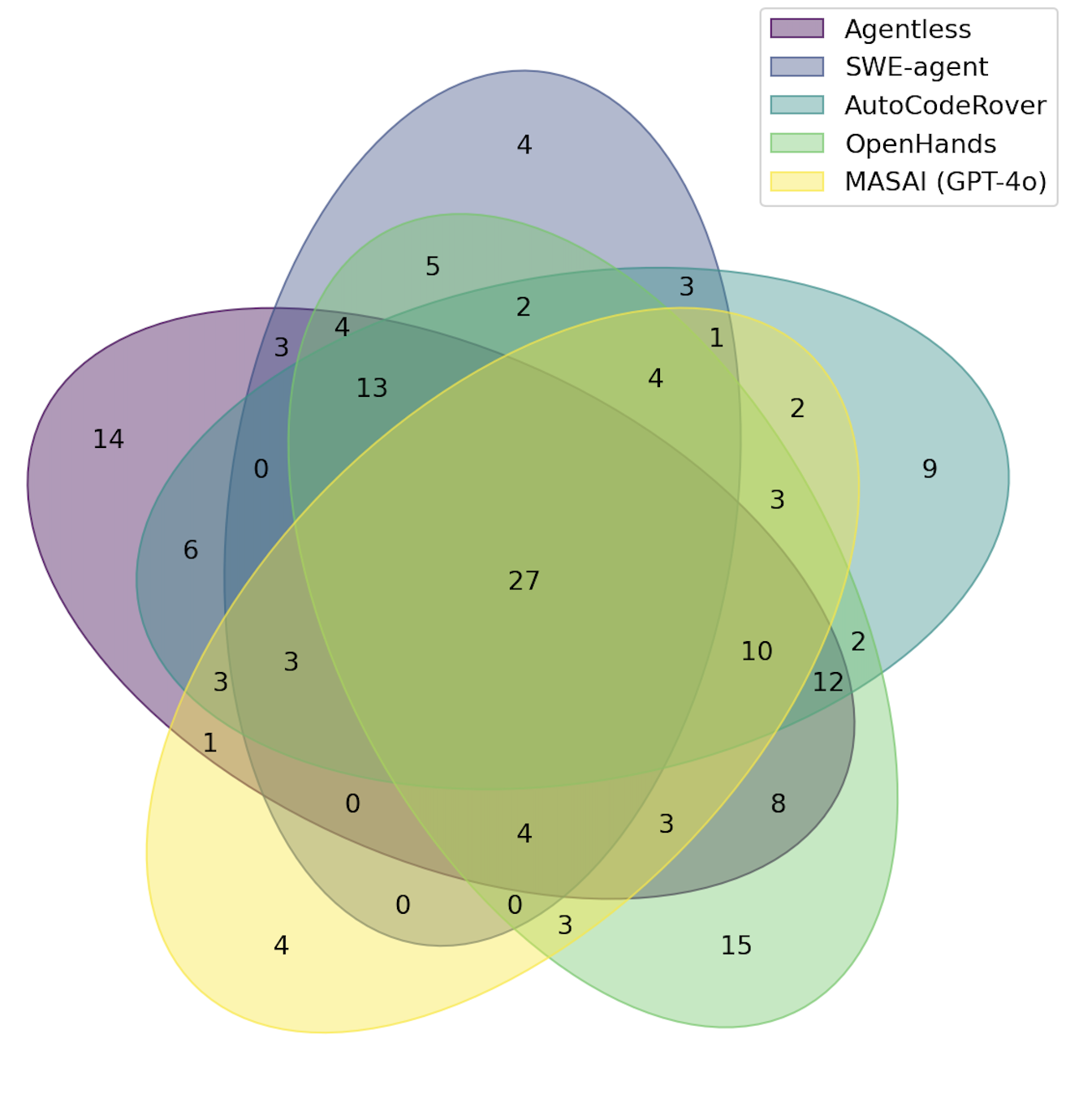}
\label{fig:venn15min-1hr}
\end{minipage}
\hspace{0.5em}
\begin{minipage}[t]{0.32\textwidth}
\centering
\scriptsize
\setlength{\tabcolsep}{2pt}
\renewcommand{\arraystretch}{1.05}
{\bfseries\small (c) Unique Solves (Moderate)}
\vspace{0.3em}

\begin{tabular}{l|p{4.0cm}}
\toprule
\textbf{Agent} & \textbf{Key Bug Types} \\
\midrule
\textbf{SA} (4) & Boundary-condition logic in rendering, callback/control-attribute propagation, type-inference precision \\
\midrule
\textbf{ACR} (9) & API naming conflicts, schema FK mapping, ORM inheritance consistency, cross-database query lookups \\
\midrule
\textbf{A} (14) & Data serialization, ORM uniqueness validation, encoding/formatting, docstring parsing (Django-heavy) \\
\midrule
\textbf{OH} (15) & Meta-programming, lazy-loading efficiency, migration deps, symbolic computation, cache handling \\
\midrule
\textbf{M} (4) & Edge-case boundaries (empty-matrix, date-parsing), metadata propagation, DataFrame indexing \\
\bottomrule
\end{tabular}
\label{tab:unique-issues}
\end{minipage}

\caption{\textbf{RQ2 Performance Analysis.} (a) Resolution rate (\%) by difficulty; \colorbox{green!60}{greener} = higher, \colorbox{red!30}{red} = 0\%. (b) Agent overlap for moderate issues; 27/168 solved by all. (c) Uniquely solved issues by agent. ACR/SA/A/OH use Claude-3.5; M uses GPT-4o.}
\label{fig:rq2-combined}
\end{figure*}

\noindent\textbf{\textsc{5.3 Results}}\quad \textbf{RQ2.1:}~Figure~\ref{fig:rq2-combined}(a) compares agent resolution rates across four difficulty tiers for GPT-4o and Claude-3.5-Sonnet backbones. Success is highest on the easiest issues (<15 minutes), showing current agentic systems handle quick, low-context fixes well. As difficulty increases, resolution drops sharply for every agent and both backbones. OH and Agentless perform especially well on the easiest tier, suggesting both structured workflows (Agentless) and flexible open-process reasoning (OH) are effective for simple problems; they also remain competitive on Moderate issues, indicating better generalization than other agents. Performance deteriorates on harder tasks: on the Hard (1--4 hour) tier, OH is best but reaches only 19\%, and most agents solve far fewer. On the most difficult issues (>4 hours), nearly all agents fail entirely (OH solves 1), highlighting current limits of LLM-based repair on highly complex bugs. Overall, the results point to a need for advances in agent reasoning, planning, and tool use to address more demanding software engineering challenges.

\textbf{RQ2.2:}~Figure~\ref{fig:rq2-combined}(b) shows that, in the Moderate tier, agents solve 168 of 261 issues, but overlap is limited: only 27 are solved by all five. Coverage differs by agent (e.g., 14 issues solved uniquely by Agentless, 15 by OH, and 9 by ACR), suggesting workflow agents (ACR, MASAI, Agentless) and open-process agents (OpenHands, SWE-Agent) succeed on different problems in ways that reflect their architectures; this pattern persists across tiers, indicating complementary strengths.\footnote{\href{https://github.com/ARiSE-Lab/understanding-apr-agents/blob/main/submission/results/T2_unique_counts.md}{Venn diagrams of resolved issues across agents and difficulty tiers.}} Manual two-reviewer annotation of uniquely solved issues (Figure~\ref{fig:rq2-combined}(c)) indicates specialization by bug type: SWE-Agent on boundary-condition logic, ACR on cross-framework API/schema issues, Agentless on serialization/ORM bugs (notably Django), OpenHands on meta-programmed framework metadata drift, and MASAI on edge-case boundaries and metadata propagation.

\textbf{RQ2.3:}~\Cref{fig:rq1-threeway}(b) further demonstrates that the availability of more issue-reproducing (i.e., fail-to-pass) test cases has a significant impact on agent patch success rates. When such test cases exist, agents face stricter constraints during patch validation, reducing the likelihood of overfitting or producing spurious patches (i.e., semantically incorrect yet plausible patches). As a result, resolving the issue according to the benchmark evaluation becomes more challenging because agents must satisfy a larger set of test constraints. This effect is consistently observed across all difficulty levels and is further corroborated by other contemporary studies~\cite{wang2025solved, ruan2024specrover}, highlighting its generality.

We assess the effect of test-case count with a group-wise t-test comparing patch success for issues with $\leq$1 fail-to-pass test versus 2--5. These groups cover 95\% of the dataset; we exclude rare outliers with exceptionally many tests (sometimes >400) to avoid skew. Although higher resolution in the $\leq$1 group could be attributed to a higher share of easier issues, Figure~\ref{fig:rq1-threeway}(c) rules this out: both groups have nearly identical difficulty distributions. We find statistically significant differences with large effect sizes—issues with >1 fail-to-pass test have substantially lower resolution rates—and this holds consistently across all agents and backbones. This suggests that additional fail-to-pass tests pose a substantial challenge for current agent-based repair and may reduce overfitting.

To measure overfitting, we analyze SWE-agent and Agentless trajectories using Claude-3.5-Sonnet and GPT-4o. We label a patch as \emph{overfitting} if it passes the agent-generated reproducer test but either (i) fails the original \texttt{FAIL\_TO\_PASS} test(s) or (ii) breaks previously passing tests in the \texttt{PASS\_TO\_PASS} suite; the overfitting rate is the fraction of patched tasks that pass the reproducer but fail the ground-truth tests. For Agentless, we programmatically extract designated marker strings from structured trajectory logs (e.g., ``Issue Reproduced''/``Issue Resolved'') to identify pre-/post-patch checkpoints and compare them against the \texttt{FAIL\_TO\_PASS} and \texttt{PASS\_TO\_PASS} suites; for SWE-agent, we similarly extract reproduction and verification flags from its logs. Figure~\ref{fig:rq1-threeway}(a) shows overfitting in both SWE-agent (17--26\%) and Agentless (4--5\%), implying that reproducer-passing patches may not generalize beyond the reproducer, and Agentless's lower rate suggests structured, workflow-based validation yields more reliable testing-phase patch assessment.

\begin{findingbox}
\textbf{RQ2 Findings:} Agent success drops from 66--73\% on easy tasks to 4.8--19\% on hard tasks, with none succeeding on very hard issues. Only 27 of 261 moderate issues are solved by all agents, with each uniquely solving 4--15, suggesting ensemble potential. More fail-to-pass tests reduce spurious patches; overfitting is higher in open-process agents (17--26\%) than workflow (4--5\%).

\end{findingbox}

\begin{figure}[!t]
\centering

\begin{minipage}[t]{0.30\textwidth} % Left plot (a): Overfitting rate
\centering
\begin{tikzpicture}
\begin{axis}[
    ybar,
    width=1.15\linewidth,
    height=3.6cm,
    bar width=5pt,
    ymin=0, ymax=40,
    enlarge x limits=0.2,
    xticklabel style={rotate=45,anchor=east, font=\small},
    ylabel={Overfitting Rate (\%)},
    y label style={font=\normalsize},
    yticklabel style={font=\normalsize},
    symbolic x coords={Easy, Moderate, Hard, All},
    xtick=data,
    legend style={
        at={(0.5,-0.58)},
        anchor=north,
        draw=none,
        legend columns=4,
        font=\scriptsize,
        column sep=2pt
    },
    yticklabel pos=left,
]
% SWE-agent Claude 3.5
\addplot+[fill=blue!70,bar shift=-7.5pt] coordinates {
    (Easy,20.6) (Moderate,28.9) (Hard,34.1) (All,26.3)
};
% SWE-agent GPT-4o
\addplot+[fill=blue!30,bar shift=-2.5pt] coordinates {
    (Easy,17.1) (Moderate,18.3) (Hard,12.5) (All,17.2)
};
% Agentless Claude 3.5
\addplot+[fill=orange!70,bar shift=2.5pt] coordinates {
    (Easy,2.1) (Moderate,5.0) (Hard,2.4) (All,3.6)
};
% Agentless GPT-4o
\addplot+[fill=orange!30,bar shift=7.5pt] coordinates {
    (Easy,2.1) (Moderate,6.2) (Hard,7.1) (All,4.9)
};
\legend{SA-C, SA-G, A-C, A-G}
\end{axis}
\end{tikzpicture}
\caption*{\small \textbf{(a)} Overfitting rate}
\end{minipage}%
\hfill
\begin{minipage}[t]{0.32\textwidth} % Middle plot (b): Patch success
\raggedright
\begin{tikzpicture}
\begin{axis}[
    ybar,
    width=1.16\linewidth,
    height=3.6cm,
    xticklabel style={rotate=45,anchor=east, font=\small},
    bar width=4pt,
    ymin=0, ymax=70,
    enlarge x limits=0.06,
    ylabel={Resolution Rate (\%)},
    y label style={font=\normalsize},
    yticklabel style={font=\normalsize},
    symbolic x coords={OH-C,OH-G,ACR-C,A-C,A-G,M-G,SA-C,SA-G},
    legend style={
        at={(0.5,-0.55)},
        anchor=north,
        draw=none,
        legend columns=2,
        font=\scriptsize
    },
    xtick=data,
    yticklabel pos=left,
]
\addplot+[fill=blue!80,bar shift=-2.5pt] coordinates {
    (OH-C,60.1) (OH-G,31.4) (ACR-C,54.4) (A-C,48.4) (A-G,42.8) (M-G,39.4) (SA-C,38.0) (SA-G,28.5)
};
\addplot+[fill=orange!90,bar shift=+2.5pt] coordinates {
    (OH-C,39.2) (OH-G,15.5) (ACR-C,29.2) (A-C,30.3) (A-G,23.7) (M-G,20.5) (SA-C,23.9) (SA-G,19.9)
};
\legend{F2P Tests $\leq 1$, F2P Tests $> 1$}
\end{axis}
\end{tikzpicture}
\caption*{\small \textbf{(b)} Pass rate by \# Fail\_to\_Pass tests}
\end{minipage}%
\hfill
\begin{minipage}[t]{0.30\textwidth} % Right plot (c): Solve rate comparison
\centering
\begin{tikzpicture}
\begin{axis}[
    width=1.1\linewidth,
    height=3.6cm,
    ylabel={Proportion (\%)},
    y label style={font=\normalsize},
    xticklabel style={rotate=45,anchor=east, font=\small},
    symbolic x coords={Easy,Moderate,Hard},
    xtick=data,
    ymin=0, ymax=70,
    ymajorgrids=true,
    yticklabel style={font=\normalsize},
    legend style={
        at={(0.5,-0.55)},
        anchor=north,
        draw=none,
        legend columns=2,
        font=\scriptsize
    },
    legend cell align={left},
    mark options={scale=1.0},
    every axis plot/.append style={thick},
    yticklabel pos=left,
]
\addplot[color=blue, mark=*] coordinates {(Easy,40.52) (Moderate,54.81) (Hard,4.66)};
\addplot[color=orange, mark=square*] coordinates {(Easy,35.71) (Moderate,47.40) (Hard,16.88)};
\legend{$\leq 1$ test,$>1$ tests}
\end{axis}
\end{tikzpicture}
\caption*{\small \textbf{(c)} Issue difficulty distribution}
\end{minipage}

\caption{\textbf{Fail-to-Pass (F2P) test analysis.} (a) Overfitting rate by difficulty for SWE-agent (SA) and Agentless (A) with Claude-3.5 (C) and GPT4o (G). (b) Issue pass rate by number of F2P tests for OpenHands (OH), AutoCodeRover (ACR), Agentless, MASAI (M), SWE-agent. (c) Distribution of task difficulty across test counts.}
\label{fig:rq1-threeway}
\end{figure}

\newpage
\section{Agent Effectiveness in Validating Patches through Testing (RQ3)}
\label{sec:rq2}

\noindent\textbf{\textsc{6.1 Reproduction Testing}}

\noindent\textbf{\textsc{6.1.1 Motivation}}\quad
RQ2 showed that fail-to-pass tests prevent agents from generating incorrect but plausible patches by providing clear failure signals that enforce semantic correctness. However, such tests are often missing when a bug is reported, requiring agents to generate them during repair. This motivates us to study how effectively agents generate issue-reproducing test cases. We explore how often agent-generated reproducers trigger bugs, pass with agent-generated patches, and lead to task resolution. We then experiment with design choices informed by our taxonomy (Section~2.2), as in RQ3.2, which explores the impact of bug localization on test generation.

\vspace{0.4em}
\noindent\textbf{\textsc{6.1.2 Approach}}\quad
To identify agent-generated reproducers, we analyze each trajectory step and automatically extract any newly created Python files the agent executes. Workflow agents follow explicitly formatted, pre-defined workflows, enabling direct parsing of generated reproducers; for open-process agents, we treat any newly created Python file the agent runs as a reproducer, even if it fails to trigger the bug. Manual trajectory inspection confirms these files are reproducer tests generated during the ``Issue Reproduction'' phase of our taxonomy, and we compute reproducer-generation rates across all agent--LLM combinations. To determine whether reproducers trigger the bug and are later resolved by a patch, we focus on Agentless\footnote{All Agentless RQ3 trajs sourced from ByteDance-Seed~\cite{multiswebenchtrajs} due to format requirements in our pipeline.} and SWE-Agent: Agentless prints designated marker strings (e.g., ``Issue reproduced'', ``Issue resolved'') when the buggy behavior is observed or fixed, so we detect reproduction by checking for these markers in execution output, while for SWE-Agent we extract analogous reproduction/resolution flags from its trajectory logs.

\definecolor{darkpurple}{RGB}{106,50,159}

\begin{figure*}[!t]
    \centering
    \hspace{-0.6em}
    \begin{minipage}[t]{0.38\textwidth}
    \centering
    {\small\bfseries (a) Issue Resolution}
    \vspace{0.1em}

    \begin{tikzpicture}
    \begin{groupplot}[
        group style={
            group size=2 by 2,
            /pgf/bar width=4pt,
            horizontal sep=0.4cm,
            vertical sep=1.0cm,
            ylabels at=edge left,
        },
        width=0.64\textwidth,
        height=0.53\textwidth,
        ymin=0, ymax=100,
        ybar stacked,
        enlarge x limits=0.15,
        ylabel={\% Tasks},
        ylabel shift=-7pt,
        ylabel style={font=\scriptsize},
        symbolic x coords={SA-C, SA-G, OH-C, OH-G, A-C, A-G, ACR-C, M-G},
        xtick=data,
        xticklabel style={rotate=50, anchor=east, font=\tiny},
        yticklabel style={font=\scriptsize},
        legend style={at={(0.0,-0.45)}, anchor=north, legend columns=1, font=\tiny, column sep=3pt, draw=none}
    ]

    \nextgroupplot[title={All (N=500)},title style={font=\scriptsize,yshift=-7pt}]
    \addplot[blue!70,fill=blue!70] coordinates {
        (SA-C,23.2) (OH-C,52.8) (A-C,42.4) (ACR-C, 46.2) (SA-G, 20.8) (OH-G, 20.0) (A-G, 36.2) (M-G, 32.6)
    };
    \addplot[red!80,fill=red!80] coordinates {
        (SA-C,46.6) (OH-C,45.4) (A-C,57.4) (ACR-C, 53.8) (SA-G, 59.4) (OH-G, 51.8) (A-G, 63.8) (M-G, 67.4)
    };

    \nextgroupplot[title={Easy (N=194)},title style={font=\scriptsize,yshift=-7pt}]
    \addplot[blue!70,fill=blue!70] coordinates {
        (SA-C,31.4) (OH-C,72.2) (A-C,61.9) (ACR-C, 66.0) (SA-G, 33.5) (OH-G, 30.4) (A-G, 55.2) (M-G, 47.9)
    };
    \addplot[red!80,fill=red!80] coordinates {
        (SA-C,36.1) (OH-C,26.8) (A-C,38.1) (ACR-C, 34.0) (SA-G, 43.3) (OH-G, 40.7) (A-G, 44.8) (M-G, 52.1)
    };

    \nextgroupplot[title={Moderate (N=261)},title style={font=\scriptsize,yshift=-7pt}]
    \addplot[blue!70,fill=blue!70] coordinates {
        (SA-C,20.3) (OH-C,44.1) (A-C,34.1) (ACR-C, 38.3) (SA-G, 14.6) (OH-G, 14.9) (A-G, 28.0) (M-G, 26.1)
    };
    \addplot[red!80,fill=red!80] coordinates {
        (SA-C,51.7) (OH-C,54.4) (A-C,65.5) (ACR-C, 61.7) (SA-G, 67.8) (OH-G, 57.9) (A-G, 72.0) (M-G, 73.9)
    };

    \nextgroupplot[title={Hard (N=42)},title style={font=\scriptsize,yshift=-7pt}]
    \addplot[blue!70,fill=blue!70] coordinates {
        (SA-C,4.8) (OH-C,19.1) (A-C,7.1) (ACR-C, 7.1) (SA-G, 2.4) (OH-G, 4.8) (A-G, 2.4) (M-G, 4.8)
    };
    \addplot[red!80,fill=red!80] coordinates {
        (SA-C,59.5) (OH-C,73.8) (A-C,92.9) (ACR-C, 92.9) (SA-G, 81.0) (OH-G, 66.7) (A-G, 97.6) (M-G, 95.2)
    };

    \legend{Resolved w/ Reproducer, Unresolved w/ Reproducer}

    \end{groupplot}
    \end{tikzpicture}
    \label{fig:reproducer-resolved}
    \end{minipage}
    \hspace{0.2em}
    \begin{minipage}[t]{0.18\textwidth}
    \centering
    {\small\bfseries\mbox{(b) Reproducer Analysis}}
    \vspace{0.2em}

    \tiny
    \setlength{\tabcolsep}{1pt}
    \renewcommand{\arraystretch}{0.95}
    \begin{tabular}{c|c|rrrrr}
    \toprule
    \textbf{D} & & \textbf{Gen.} & \textbf{Trig.} & \textbf{Res.} & \textbf{Ofit.} & \textbf{Solv.} \\
    \midrule
    \multirow{4}{*}{\rotatebox{90}{\textbf{Easy}}}
    & A-C & \cellcolor{darkpurple}\textcolor{white}{99.0} & 41.8 & 26.8 & 2.1 & 62.5 \\
    & \cellcolor{gray!10}A-G & \cellcolor{blue!48}96.9 & \cellcolor{gray!10}40.2 & \cellcolor{gray!10}22.2 & \cellcolor{gray!10}2.1 & \cellcolor{gray!10}56.9 \\
    & SA-C & \cellcolor{blue!38}76.7 & 49.2 & 50.8 & \cellcolor{red!20}24.9 & 49.2 \\
    & \cellcolor{gray!10}SA-G & \cellcolor{blue!46}92.6 & \cellcolor{gray!10}54.3 & \cellcolor{gray!10}39.4 & \cellcolor{gray!10}17.7 & \cellcolor{gray!10}40.6 \\
    \midrule
    \multirow{4}{*}{\rotatebox{90}{\textbf{Mod.}}}
    & A-C & \cellcolor{darkpurple}\textcolor{white}{98.9} & 41.8 & 18.4 & 5.0 & 34.5 \\
    & \cellcolor{gray!10}A-G & \cellcolor{darkpurple}\textcolor{white}{98.9} & \cellcolor{gray!10}45.2 & \cellcolor{gray!10}15.7 & \cellcolor{gray!10}6.1 & \cellcolor{gray!10}28.3 \\
    & SA-C & \cellcolor{blue!43}85.4 & 62.8 & 53.0 & \cellcolor{red!20}32.0 & 28.9 \\
    & \cellcolor{gray!10}SA-G & \cellcolor{blue!46}93.0 & \cellcolor{gray!10}55.9 & \cellcolor{gray!10}31.0 & \cellcolor{gray!10}19.2 & \cellcolor{gray!10}19.2 \\
    \midrule
    \multirow{4}{*}{\rotatebox{90}{\textbf{Hard}}}
    & A-C & \cellcolor{darkpurple}\textcolor{white}{100.0} & 35.7 & 4.8 & 2.4 & 7.1 \\
    & \cellcolor{gray!10}A-G & \cellcolor{darkpurple}\textcolor{white}{100.0} & \cellcolor{gray!10}38.1 & \cellcolor{gray!10}7.1 & \cellcolor{gray!10}7.1 & \cellcolor{gray!10}2.4 \\
    & SA-C & \cellcolor{blue!41}82.9 & 58.5 & 39.0 & \cellcolor{red!25}36.6 & 4.9 \\
    & \cellcolor{gray!10}SA-G & \cellcolor{blue!46}92.5 & \cellcolor{gray!10}55.0 & \cellcolor{gray!10}12.5 & \cellcolor{gray!10}12.5 & \cellcolor{gray!10}2.5 \\
    \midrule
    \multirow{4}{*}{\rotatebox{90}{\textbf{All}}}
    & A-C & \cellcolor{darkpurple}\textcolor{white}{99.0} & 41.2 & 20.4 & 3.6 & 42.8 \\
    & \cellcolor{gray!10}A-G & \cellcolor{darkpurple}\textcolor{white}{98.2} & \cellcolor{gray!10}42.6 & \cellcolor{gray!10}17.6 & \cellcolor{gray!10}4.8 & \cellcolor{gray!10}36.9 \\
    & SA-C & \cellcolor{blue!41}81.9 & 57.4 & 51.0 & \cellcolor{red!20}29.8 & 34.6 \\
    & \cellcolor{gray!10}SA-G & \cellcolor{blue!46}92.8 & \cellcolor{gray!10}55.3 & \cellcolor{gray!10}32.4 & \cellcolor{gray!10}17.9 & \cellcolor{gray!10}26.0 \\
    \bottomrule
    \end{tabular}
    \label{tab:reproducer-counts}
    \end{minipage}
    \hfill
    \begin{minipage}[t]{0.38\textwidth}
    \centering
    {\small\bfseries (c) Regression Test Selection}
    \vspace{0.1em}

    \begin{tikzpicture}
    \begin{groupplot}[
        group style={
            group size=2 by 2,
            /pgf/bar width=4pt,
            horizontal sep=0.4cm,
            vertical sep=1.0cm,
            ylabels at=edge left,
        },
        width=0.64\textwidth,
        height=0.53\textwidth,
        ymin=0, ymax=100,
        ybar stacked,
        enlarge x limits=0.15,
        ylabel={\% Tasks},
        ylabel shift=-7pt,
        ylabel style={font=\scriptsize},
        symbolic x coords={SA-C, SA-G, OH-C, OH-G, A-C, A-G, ACR-C, M-G},
        xtick=data,
        xticklabel style={rotate=50, anchor=east, font=\tiny},
        yticklabel style={font=\scriptsize},
        legend style={at={(-0.15,-0.45)}, anchor=north, legend columns=2, font=\tiny, column sep=2pt, draw=none}
    ]

    \nextgroupplot[title={All (N=500)},title style={font=\scriptsize,yshift=-7pt}]
    \addplot[yellow!80!orange,fill=yellow!80!orange] coordinates {
        (SA-C,6.4) (OH-C,0.2) (A-C,26.0) (ACR-C, 53.0) (SA-G, 3.2) (OH-G, 5.4) (A-G, 27.0) (M-G, 3.2)
    };
    \addplot[yellow!80!orange,fill=yellow!80!orange, postaction={pattern=north east lines, pattern color=orange}] coordinates {
        (SA-C,6.0) (OH-C,0.2) (A-C,19.2) (ACR-C, 46.2) (SA-G, 1.2) (OH-G, 3.4) (A-G, 17.8) (M-G, 0.8)
    };
    \addplot[blue!70,fill=blue!70, postaction={pattern=north west lines, pattern color=green}] coordinates {
        (SA-C,4.6) (OH-C,0.6) (A-C,22.2) (ACR-C, 0) (SA-G, 3.6) (OH-G, 5.0) (A-G, 17.6) (M-G, 27.2)
    };
    \addplot[blue!70,fill=blue!70] coordinates {
        (SA-C,8.2) (OH-C,1.6) (A-C,30.0) (ACR-C, 0) (SA-G, 5.8) (OH-G, 10.2) (A-G, 35.0) (M-G, 54.0)
    };

    \nextgroupplot[title={Easy (N=194)},title style={font=\scriptsize,yshift=-7pt}]
    \addplot[yellow!80!orange,fill=yellow!80!orange] coordinates {
        (SA-C,5.7) (OH-C,0.0) (A-C,19.6) (ACR-C, 33.0) (SA-G, 3.1) (OH-G, 5.2) (A-G, 17.5) (M-G, 2.6)
    };
    \addplot[yellow!80!orange,fill=yellow!80!orange, postaction={pattern=north east lines, pattern color=orange}] coordinates {
        (SA-C,8.2) (OH-C,0.2) (A-C,26.3) (ACR-C, 66.0) (SA-G, 3.1) (OH-G, 5.7) (A-G, 27.8) (M-G, 1.0)
    };
    \addplot[blue!70,fill=blue!70, postaction={pattern=north west lines, pattern color=green}] coordinates {
        (SA-C,4.6) (OH-C,1.0) (A-C,33.5) (ACR-C, 0) (SA-G, 5.7) (OH-G, 6.7) (A-G, 25.8) (M-G, 40.7)
    };
    \addplot[blue!70,fill=blue!70] coordinates {
        (SA-C,5.2) (OH-C,0.5) (A-C,18.0) (ACR-C, 0) (SA-G, 5.2) (OH-G, 7.7) (A-G, 25.8) (M-G, 42.3)
    };

    \nextgroupplot[title={Moderate (N=261)},title style={font=\scriptsize,yshift=-7pt}]
    \addplot[yellow!80!orange,fill=yellow!80!orange] coordinates {
        (SA-C,6.9) (OH-C,0.4) (A-C,27.6) (ACR-C, 60.9) (SA-G, 2.7) (OH-G, 5.7) (A-G, 30.7) (M-G, 3.4)
    };
    \addplot[yellow!80!orange,fill=yellow!80!orange, postaction={pattern=north east lines, pattern color=orange}] coordinates {
        (SA-C,5.4) (OH-C,0.0) (A-C,16.5) (ACR-C, 38.3) (SA-G, 0.0) (OH-G, 2.3) (A-G, 13.0) (M-G, 0.8)
    };
    \addplot[blue!70,fill=blue!70, postaction={pattern=north west lines, pattern color=green}] coordinates {
        (SA-C,5.4) (OH-C,0.4) (A-C,17.2) (ACR-C, 0) (SA-G, 2.7) (OH-G, 4.6) (A-G, 14.6) (M-G, 21.5)
    };
    \addplot[blue!70,fill=blue!70] coordinates {
        (SA-C,7.3) (OH-C,1.9) (A-C,36.4) (ACR-C, 0) (SA-G, 6.1) (OH-G, 11.9) (A-G, 39.8) (M-G, 58.6)
    };

    \nextgroupplot[title={Hard (N=42)},title style={font=\scriptsize,yshift=-7pt}]
    \addplot[yellow!80!orange,fill=yellow!80!orange] coordinates {
        (SA-C,7.1) (OH-C,0.0) (A-C,47.6) (ACR-C, 92.9) (SA-G, 7.1) (OH-G, 4.8) (A-G, 50.0) (M-G, 2.4)
    };
    \addplot[yellow!80!orange,fill=yellow!80!orange, postaction={pattern=north east lines, pattern color=orange}] coordinates {
        (SA-C,0.0) (OH-C,0.0) (A-C,4.8) (ACR-C, 7.1) (SA-G, 0.0) (OH-G, 0.0) (A-G, 2.4) (M-G, 0.0)
    };
    \addplot[blue!70,fill=blue!70, postaction={pattern=north west lines, pattern color=green}] coordinates {
        (SA-C,0.0) (OH-C,0.0) (A-C,2.4) (ACR-C, 0) (SA-G, 0.0) (OH-G, 0.0) (A-G, 0.0) (M-G, 2.4)
    };
    \addplot[blue!70,fill=blue!70] coordinates {
        (SA-C,28.6) (OH-C,4.8) (A-C,40.5) (ACR-C, 0) (SA-G, 7.1) (OH-G, 11.9) (A-G, 42.9) (M-G, 78.6)
    };

    \legend{Gold (Unres.), Gold (Res.), Non-Gold (Res.), Non-Gold (Unres.)}

    \end{groupplot}
    \end{tikzpicture}
    \label{fig:regression-all-v-gold-V2}
    \end{minipage}
    \hspace{-0.2em}

    \caption{\textbf{Reproducer and Regression Testing.} \textbf{(a)} Issue resolution of tasks where agents generate a reproducer: resolved (blue) vs unresolved (red). \textbf{(b)} Reproducer analysis (\%) by difficulty (D) computed over tasks with a generated patch: Gen.=generates a reproducer, Trig.=triggers the bug, Res.=patch passes reproducer, Ofit.=patch passes reproducer but fails task, Solv.=task resolved. Darker shading indicates higher values; for Ofit., red shading indicates more overfitting. \textbf{(c)} Regression test selection: tasks where agents run gold tests (yellow) vs non-gold tests (blue). Striped=resolved, solid=unresolved. Agents: SA=SWE-Agent, OH=OpenHands, A=Agentless, ACR=AutoCodeRover, M=MASAI, with -C (Claude-3.5) or -G (GPT-4o).}
    \label{fig:reproducer-regression-combined}
\end{figure*}

\vspace{0.4em}
\noindent\textbf{\textsc{6.1.3 Analysis of Agent Trajectories}}\quad
We first measure reproducer test-generation frequency and task resolution rates. Figure~\ref{fig:reproducer-regression-combined}(a) shows workflow agents Agentless, ACR, and MASAI generate reproducers for all tasks, no matter the underlying LLM. Open-process agents generate reproducers less often (<80\% of tasks), except for OpenHands with Claude-Sonnet-3.5. This difference likely emerges because workflow agents often have predefined workflow steps dedicated to issue reproduction while open-process agents may generate a patch without attempting to reproduce the issue.
For all agents, task resolution rates decrease with difficulty regardless of reproducer count, motivating deeper analysis of whether generated reproducers successfully trigger bugs.

We analyze reproducer effectiveness in Agentless and SWE-Agent. As shown in Figure~\ref{fig:reproducer-regression-combined}(b), Agentless generates reproducers for nearly all tasks (98--99\%) while SWE-Agent ranges from 82--93\%. Trigger rates are 41--57\%, with Agentless showing lower overfitting (4--5\%) than SWE-Agent (18--30\%). Agentless tends to have lower reproducer resolution rates than SWE-Agent, but significantly lower overfitting rates. SWE-Agent maintains higher resolution rates, but many reproducers overfit, so tasks are not ultimately resolved. This suggests Agentless produces effective reproducers but struggles to generate valid fixes, while the disparity likely stems from Agentless' structured workflow steps for reproduction versus SWE-Agent's open-process design. The relatively low triggering rate ($\sim$40\%) motivates further strategies for improving reproducer generation.

\vspace{0.5em}
\noindent\textbf{\textsc{6.1.4 Improving Reproducer Generation}}\quad The low triggering rate ($\sim$40\%) motivates strategies to improve reproducer generation. We study two factors: (1) bug localization context, an early-pipeline signal that can improve test generation; and (2) inference scaling (sampling and temperature), challenging the assumption that scaling universally improves performance by investigating where it helps and where diminishing returns emerge. Both are hyperparameters and contextual signals easily integrable into existing pipelines. We define our \textbf{problem statement} as follows: given a codebase $C_o$ and bug report $b$ (natural language, possibly with stack traces), synthesize a test $T_{\text{gen}}$ (e.g., a \texttt{.py} file) that triggers the bug on $C_o$. Agents formulate a prompt $P(b)$ with the issue description, task instructions, and often an example test, then pass it to the LLM. Some agents (e.g., Agentless) generate multiple samples in parallel (e.g., 40 at once). Test generation succeeds if at least one test reproduces the bug, and successful tests are reused for patch validation.

\textbf{Experimental Setup.} We use Agentless on Verified-50, as it is the best-performing workflow agent, enabling easier reproduction when varying components. For each issue, we retrieve the repository at base commit ($C_o$), build a Docker image with the needed environment, and execute $T_{\text{gen}}$ in containers. Workers are capped for runtime stability. All experiments use Claude Sonnet 4 with safeguards to avoid rate-limit errors: limiting output tokens, batching prompts within token budgets, and throttling requests with 60-second intervals. In the \textit{sampling and temperature} experiment, we vary sample count $s \in \{15, 40\}$ and temperature $t \in \{0.2, 0.6, 1.0\}$ (lower $t$ = more deterministic, higher = more diverse), sampling $s$ tests for each $(s, t)$ setting and evaluating reproduction success as defined above. In the \textit{bug localization} experiment, we generate 30 $T_{\text{gen}}$ samples per issue and compare two configurations: (1) \textit{Default} and (2) \textit{with bug localization}, which enhances the prompt with localization context (relevant files, classes, functions, lines) extracted by embedding-based retrieval. Success is measured as $k/30$ where $k$ is the count of samples that reproduce the issue.

\textbf{RQ3.1~Inference time scaling using sampling with different temperature.} We evaluate all 50 issues across sample counts $s \in \{15, 40\}$ and temperatures $t \in \{0.2, 0.6, 1.0\}$.\footnote{\href{https://github.com/ARiSE-Lab/understanding-apr-agents/blob/main/submission/results/samp_vs_temp.pdf}{Verified-50 sampling vs temperature.}} Low temperature ($t = 0.2$) outperforms higher temperatures: success rates are 46.0\% vs.\ 42--43\% across both sample sizes (see \Cref{tab:rq3-summary}). Increased sampling yields more successful tests per issue but does not necessarily increase the proportion of issues reproduced; the optimal strategy is more samples at low temperature rather than high-temperature diversity. 11 issues (22\%) fail across all configurations due to vague descriptions, while 5 (10\%) achieve near-perfect success regardless of settings.

\noindent\textbf{Error Analysis.} When test generation repeatedly fails, we analyze why and quantify how failures vary with temperature. Common modes include import/configuration errors, missing migrations, and model mismatches. For each 40-sample run, we extract error patterns from execution logs and classify each output by pattern matching as \texttt{SUCCESS} (issue reproduced), \texttt{NOT\_REPRODUCED} (test failed to trigger the bug), or an error type (e.g., \texttt{IMPORT\_ERROR}, \texttt{ATTRIBUTE\_ERROR}, \texttt{FILE\_NOT\_FOUND}, \texttt{TYPE\_ERROR}, \texttt{SYNTAX\_ERROR}, \texttt{TEST\_EXECUTION\_FAILED}). We measure diversity as the number of unique error types per issue at each $t$.\footnote{\href{https://github.com/ARiSE-Lab/understanding-apr-agents/blob/main/submission/results/T9_error_diversity.md}{Verified-50 error diversity analysis.}} Low temperature ($t$=0.2) yields repetitive failures, while higher temperatures surface more diverse failures: average unique error types increase from 2.08 ($t$=0.2) to 2.44 ($t$=0.6) to 2.76 ($t$=1.0), a 32.6\% gain. However, success rates decline with temperature, revealing a trade-off: higher temperatures broaden the search space but mostly discover new failure modes rather than solutions (e.g., \textit{django-11265} drops from 82.5\% success at $t$=0.2 to 50.0\% at $t$=1.0 and unique errors rise from 2 to 5; \Cref{tab:rq3-summary}(c)). Of Verified-50 issues, 40.8\% are stable across temperatures, 22.4\% exhibit the trade-off (diversity up, success down), and only 10.2\% improve both; for issues that fail at all temperatures, temperature changes failure modes but not outcomes. For underspecified bug reports, failures are identical across $(s, t)$ settings, suggesting inference-time scaling provides little benefit because the model repeats the same failed logic; the best strategy is maximizing samples at low temperature rather than relying on high-temperature diversity.

\begin{table*}[!t]
\caption{\textbf{RQ3 Reproduction Test Results Summary.} All experiments run on Verified-50. (a) Temperature and sampling effects on reproduction success. (b) Impact of bug localization context. (c) Error diversity trade-off illustrated for one issue; averages computed across all 50 issues. See text footnotes for full per-issue results.}
\label{tab:rq3-summary}
\centering
\footnotesize
\setlength{\tabcolsep}{3pt}
\begin{minipage}[t]{0.32\linewidth}
\centering
\textbf{(a) Temp. \& Sampling ($s{=}40$)}\\[2pt]
\rowcolors{2}{gray!10}{white}
\begin{tabular}{l|ccc}
\toprule
 & $t{=}0.2$ & $t{=}0.6$ & $t{=}1.0$ \\
\midrule
django-11265 & 33/40 & 21/40 & 20/40 \\
sympy-17139 & 40/40 & 40/40 & 40/40 \\
django-16642 & 0/40 & 0/40 & 0/40 \\
\midrule
\rowcolor{white}
\textbf{Overall} & \textbf{46.0\%} & 42.7\% & 42.3\% \\
\bottomrule
\end{tabular}
\end{minipage}\hfill\hspace{1em}
\begin{minipage}[t]{0.32\linewidth}
\centering
\textbf{(b) Bug Localization ($s{=}30$)}\\[2pt]
\rowcolors{2}{gray!10}{white}
\begin{tabular}{l|cc}
\toprule
 & Default & w/ Loc. \\
\midrule
sympy-14531 & 1/30 & 30/30 \textcolor{green!60!black}{$\uparrow$} \\
xarray-6461 & 0/30 & 5/30 \textcolor{green!60!black}{$\uparrow$} \\
django-11880 & 17/30 & 9/30 \textcolor{red}{$\downarrow$} \\
\midrule
\rowcolor{white}
\textbf{Overall} & 614 & 697 \textcolor{green!60!black}{$\uparrow$} \\
\bottomrule
\end{tabular}
\end{minipage}\hfill
\begin{minipage}[t]{0.32\linewidth}
\centering
\textbf{(c) Error Diversity ($s{=}40$)}\\[2pt]
\rowcolors{2}{gray!10}{white}
\begin{tabular}{l|ccc}
\toprule
 & $t{=}0.2$ & $t{=}0.6$ & $t{=}1.0$ \\
\midrule
Success* & 82.5\% & 52.5\% & 50.0\% \\
Unique Errors* & 2 & 4 & 5 \\
\midrule
\rowcolor{white}
\textbf{Avg. Errors} & 2.08 & 2.44 & 2.76 \\
\bottomrule
\multicolumn{4}{l}{\scriptsize *django-11265 example}
\end{tabular}
\end{minipage}
\end{table*}

\textbf{RQ3.2~Inference time scaling with Bug Localization.} Enabling bug localization significantly improves test-generation.\footnote{\href{https://github.com/ARiSE-Lab/understanding-apr-agents/blob/main/submission/results/localization_full.pdf}{Verified-50 bug localization results.}} On Verified-50, localization increases reproduction success in 54\% of issues (27/50), raising successful reproductions from 614 to 697 of 1,500 samples (13.5\% relative). It converts seven previously unreproduced issues (0/30) to reproduced ones, with some improving dramatically (1/30 to 30/30). Localization does not harm stable cases: eight issues remain at 30/30 with or without localization, and only three show minor declines. Each test requires costly API calls and execution; localization improves odds of early success, boosting speed and resource efficiency.

\begin{findingbox}
\textbf{RQ3 (Part 1) Findings.} Workflow-based agents produce reproducers more reliably with lower overfitting, but even the best fails to trigger issues 40--60\% of the time. More samples at low temperature ($t$=0.2) improves trigger rates; bug localization context helps 54\% of issues.
\end{findingbox}

\noindent\textbf{\textsc{6.2 Regression Test Selection}}

\noindent\textbf{\textsc{6.2.1 Motivation}}\quad Unlike issue-reproducing tests, large projects typically have \textit{regression tests} validating existing functionality. To resolve a bug, patched code must fix the issue and pass the test suite to preserve existing functionality. Here we explore how often APR agents run regression tests and their test selection accuracy---how often selected regression tests are relevant to the issue.
We perform this analysis for all our combinations of agents and LLM backbones across task difficulties. 

\vspace{0.5em}
\noindent\textbf{\textsc{6.2.2 Approach}}\quad
To identify regression tests, we extract Python filenames run throughout trajectories. For workflow agents, explicitly formatted trajectories enable automatic parsing of agent-run tests. For open-process agents, we consider any execution of a pre-existing repository file as a regression test. We define \textit{gold} (relevant) regression tests as the pass-to-pass tests in the SWE-bench evaluation; other project tests are \textit{non-gold}. Analysis is done at the file level: we compile files containing gold regression tests and check whether agent-executed filenames match this list.

\vspace{0.5em}

\noindent\textbf{\textsc{6.2.3 Results}}\quad Figure~\ref{fig:reproducer-regression-combined}(c) shows workflow agents run regression tests more often than open-process agents: ACR and Agentless for nearly all tasks, MASAI for >85\%. However, selection accuracy varies: MASAI runs gold tests in only 4\% of cases versus 45\% for Agentless and 100\% for ACR. This reflects design choices: Agentless runs tests upfront and an LLM filters them before patch generation; ACR runs all tests by default; MASAI lacks a dedicated step and couples regression tests with reproducer execution, accounting for its selection accuracy. Open-process agents use exploratory heuristics during or after patching, reducing the likelihood of identifying relevant tests.

Next, we examine the correlation between running regression tests and issue resolution. For each difficulty level, we find how many tasks for which agents run regression tests are subsequently resolved. Figure~\ref{fig:reproducer-regression-combined}(c) shows results where striped columns indicate resolved tasks and solid indicates unresolved. As difficulty increases, task resolution decreases (striped portions shrink), even when regression tests are run. This is expected with non-gold tests since these irrelevant tests may pass even when patches break other functionality. The trend persists with gold tests; this is unsurprising since harder tasks require more complex solutions, making resolution less likely even if existing functionality is preserved. Additionally, we measure whether agents run regression tests; even if an agent runs all gold tests, it does not mean the agent generated a patch that passed the tests.

Comparing non-gold and gold test results in Figure~\ref{fig:reproducer-regression-combined}(c), tasks with gold tests run show higher resolution rates (more relative striping in yellow bars than blue), suggesting gold tests provide better patch validation signals. Overall, while workflow agents frequently run regression tests, agents struggle to select relevant tests. Though regression tests alone are insufficient for patch validation, selecting relevant tests is important since they are more diagnostic than irrelevant ones.

\begin{findingbox}
\textbf{RQ3 (Part 2) Findings.} Tasks where agents run at least one gold regression test show higher resolution rates than those with only non-gold tests. Workflow agents run tests 90\%+ of the time but struggle to identify relevant ones (e.g., MASAI: 4\% vs Agentless: 45\%). 

\end{findingbox}

\usepgfplotslibrary{groupplots}
\section{Agent Tooling (RQ4)}

\begin{figure*}[!t]
\centering
\setlength{\tabcolsep}{0pt}
\hspace{-2.5em}
\begin{tabular}{@{}c@{\hspace{7.2em}}c@{\hspace{-1.2em}}c@{\hspace{0.7em}}}
\begin{minipage}[t]{0.24\textwidth}
\centering
{\small\bfseries (a) Agent Tool APIs}
\vspace{0.3em}

\scriptsize
\setlength{\tabcolsep}{1.5pt}
\renewcommand{\arraystretch}{1.0}
\begin{tabular}{l|c|c|ccc}
\toprule
& & \textbf{Tax.} & \multicolumn{3}{c}{\textbf{Agent-Specific Tool APIs}} \\
\cmidrule(lr){4-6}
\textbf{Cat.} & \textbf{Desc.} & \textbf{Ph.} & \textbf{OH} & \textbf{SA} & \textbf{M} \\
\midrule
\textsc{search} & \makecell[l]{Search\\[-0.3ex]project\\[-0.3ex]repo} & \makecell{Proj.\\[-0.3ex]U.} & N/A & \makecell{\texttt{search\_file},\\[-0.3ex]\texttt{find\_file},\\[-0.3ex]\texttt{search\_dir}} & \texttt{list} \\
\cmidrule{1-6}
\textsc{local.} & \makecell[l]{Mark\\[-0.3ex]buggy\\[-0.3ex]code} & \makecell{Cntxt.\\[-0.3ex]U.} & N/A & N/A & \makecell{\texttt{edit},\\[-0.3ex]\texttt{add}} \\
\cmidrule{1-6}
\textsc{view} & \makecell[l]{View\\[-0.3ex]file} & All & \texttt{str\_repl} & \makecell{\texttt{open}, \texttt{goto},\\[-0.3ex]\texttt{scroll\_down},\\[-0.3ex]\texttt{scroll\_up}} & \texttt{read} \\
\cmidrule{1-6}
\textsc{edit} & \makecell[l]{Edit\\[-0.3ex]file} & \makecell{Patch,\\[-0.3ex]Test} & \texttt{str\_repl}  & \makecell{\texttt{create},\\[-0.3ex]\texttt{edit},\\[-0.3ex]\texttt{edit\_of\_edit}} & \texttt{write} \\
\cmidrule{1-6}
\textsc{bash} & \makecell[l]{Run\\[-0.3ex]bash} & All & \makecell{\texttt{execute}\\[-0.3ex]\texttt{\_bash}} & \texttt{bash} & \texttt{command} \\
\bottomrule
\end{tabular}
 \label{tab:agent-tool-types}
\end{minipage}
&
\begin{minipage}[t]{0.3\textwidth}
\centering
{\small\bfseries (b) Action Frequency}

\begin{tikzpicture}
\begin{axis}[
    xbar=0pt,
    bar width=4pt,
    width=\textwidth,
    height=7cm,
    enlarge y limits=0.12,
    xlabel={\% All Agent Actions},
    symbolic y coords={bash, edit, view, local, search},
    ytick=data,
    xmin=0, xmax=61,
    xtick={0,25,50},
    y tick label style={font=\scshape\small},
    x label style={font=\small},
    xticklabel style={font=\small},
    nodes near coords,
    nodes near coords style={font=\tiny, xshift=1pt, anchor=west, rounded corners=1pt, inner sep=1pt},
    xmajorgrids=true,
    xminorgrids=true,
    minor x tick num=4,
    major grid style={line width=0.3pt, draw=gray!50},
    minor grid style={line width=0.2pt, draw=gray!15},
]

\addplot[xbar, bar shift=10pt, fill=blue!70,
] coordinates {
    (0.0,search) (0.0,local) (16.7,view) (38.1,edit) (45.2,bash)
};

\addplot[xbar, bar shift=5pt, fill=violet!70,
] coordinates {
    (0.0,search) (0.0,local) (23.1,view) (37.7,edit) (39.2,bash)
};

\addplot[xbar, bar shift=0pt, fill=red!80,
] coordinates {
    (12.1,search) (0.0,local) (26.9,view) (32.5,edit) (28.6,bash)
};

\addplot[xbar, bar shift=-5pt, fill=orange!80,
] coordinates {
    (9.4,search) (0.0,local) (12.1,view) (49.6,edit) (28.4,bash)
};

\addplot[xbar, bar shift=-10pt, fill=yellow!80!orange,
] coordinates {
    (14.4,search) (9.1,local) (32.1,view) (16.9,edit) (27.6,bash)
};\end{axis}
\end{tikzpicture}
\label{fig:rq4-actions}
\end{minipage}
&
\begin{minipage}[t]{0.35\textwidth}
\centering
{\small\bfseries (c) Bash Frequency}

\begin{tikzpicture}
\begin{axis}[
    xbar=0pt,
    bar width=4pt,
    width=0.9\textwidth,
    height=7cm,
    enlarge y limits=0.12,
    xlabel={\% All Bash Commands},
    symbolic y coords={find, grep, ls, cd, run},
    ytick=data,
    xmin=0, xmax=100,
    xtick={0,25,50,75,100},
    y tick label style={font=\scshape\small},
    x label style={font=\small},
    xticklabel style={font=\small, xshift=-2pt},
    nodes near coords,
    nodes near coords style={font=\tiny, xshift=1pt, anchor=west, rounded corners=1pt, inner sep=1pt},
    xmajorgrids=true,
    xminorgrids=true,
    minor x tick num=4,
    major grid style={line width=0.3pt, draw=gray!50},
    minor grid style={line width=0.2pt, draw=gray!15},
]

\addplot[xbar, bar shift=10pt, fill=blue!70,
] coordinates {
    (33.6,run) (38.6,cd) (6.6,ls) (9.6,grep) (4.0,find)
};

\addplot[xbar, bar shift=5pt, fill=violet!70,
] coordinates {
    (70.0,run) (0.1,cd) (4.7,ls) (12.6,grep) (0.9,find)
};

\addplot[xbar, bar shift=0pt, fill=red!80,
] coordinates {
    (40.1,run) (12.4,cd) (26.9,ls) (2.3,grep) (6.8,find)
};

\addplot[xbar, bar shift=-5pt, fill=orange!80,
] coordinates {
    (60.5,run) (6.7,cd) (7.6,ls) (3.2,grep) (1.8,find)
};

\addplot[xbar, bar shift=-10pt, fill=yellow!80!orange,
] coordinates {
    (80.8,run) (6.6,cd) (0.1,ls) (5.0,grep) (0.1,find)
};
\end{axis}
\end{tikzpicture}
\label{fig:rq4-bash}
\vspace{0.5em}
\end{minipage}
\end{tabular}

\vspace{-1.3em}
\hspace{22em}
\begin{tikzpicture}
\draw[fill=blue!70] (-1.5,0) rectangle (-1.25,0.12); \node[right, font=\scriptsize] at (-1.25,0.06) {OH-C};
\draw[fill=violet!70] (-0.25,0) rectangle (0.0,0.12); \node[right, font=\scriptsize] at (0.0,0.06) {OH-G};
\draw[fill=red!80] (1.0,0) rectangle (1.25,0.12); \node[right, font=\scriptsize] at (1.25,0.06) {SA-C};
\draw[fill=orange!80] (2.25,0) rectangle (2.5,0.12); \node[right, font=\scriptsize] at (2.5,0.06) {SA-G};
\draw[fill=yellow!80!orange] (3.5,0) rectangle (3.75,0.12); \node[right, font=\scriptsize] at (3.75,0.06) {M-G};
\end{tikzpicture}

\caption{\textbf{Agent Tooling.} (a) Tool APIs by category. Cat.=Category, Desc.=Description. OH=OpenHands, SA=SWE-Agent, M=MASAI. <Agent>-C=Claude-Sonnet-3.5-backed agent, <agent>-G=GPT-4o-backed agent. Tax.\ Ph.\ indicates which taxonomy phase commonly uses a tool category (Proj.\ U.=Project Understanding, Cntxt.\ U.=Context Understanding, Patch=Patching, Test=Testing). Only tools used in SWE-bench trajectories are shown; \texttt{str\_repl}=\texttt{str\_replace\_editor}. (b) Relative frequency of action categories by agent across all tasks. (c) Top-5 bash commands across all tasks; \textsc{run}=running Python files.}
\label{fig:rq4-combined}
\end{figure*}

\noindent\textbf{\textsc{7.1 Motivation}}\quad Agent actions take the form of tool APIs invoked to perform specific functions, and these tools are often specialized for each agent, resulting in diverse action spaces. We classify such tools into common categories to compare usage across OpenHands, SWE-Agent, and MASAI with all available LLM backbones, considering only tools used on SWE-bench (which excludes OpenHands web browsing). We exclude Agentless and ACR because their workflow implementations preclude LLM action selection.

\noindent\textbf{\textsc{7.2 Results}}\quad We compile tool APIs available to each agent and categorize them into common groups for cross-agent comparison. Figure~\ref{fig:rq4-combined}(a) shows such categories and their constituent APIs, revealing that while specific tools vary among agents, overall functionalities are similar.
Moreover, agent tools implement basic functionality like viewing and editing files, not encapsulating complex high-level procedures like localization or program analysis, which differs from human workflows that often use sophisticated tools like debuggers.
Figure~\ref{fig:rq4-combined}(b) shows agents vary in tool usage. MASAI uses many specialized tools, while OpenHands has fewer and thus uses bash more frequently, regardless of LLM backbone. Figure~\ref{fig:rq4-combined}(c) shows that while bash usage frequency varies, all studied agents leverage it most often for running Python files, followed by changing directories, listing contents, and searching. MASAI uses bash almost exclusively for running Python files, more than 10 percentage points more than any other agent, likely because MASAI has specialized tools that replace common bash commands (e.g., \texttt{list} replaces \texttt{ls}). Interestingly, while action frequency is largely robust to LLM choice (Figure~\ref{fig:rq4-combined}b), the GPT-backed agent configurations when using bash tend to run Python commands more frequently and \texttt{cd} and \texttt{ls} commands less frequently compared to their Claude-Sonnet-backed counterparts (Figure~\ref{fig:rq4-combined}c).
Our findings are also reflected in the taxonomy (Figure~\ref{fig:agent-taxonomy}), which was constructed from trajectories of five agents, yet has a maximum edge weight of three; there is no step where all agents choose the same two sequential actions. Similarly, agents use tools differently, and are not equipped with the same tools, providing insight for why decision pathways differ.

\begin{findingbox}
\textbf{RQ4 Findings.} Despite their sophistication and diverse implementations, the specialized action spaces of APR agents reduce to similar basic, low-level tools, mainly consisting of interfaces for viewing or editing files and executing arbitrary bash commands. 
\end{findingbox}

\section{Related Work}

\textbf{SWE Agents for Program Repair.} AI-assisted tools are increasingly adopted in industry~\cite{klemmer2024using, vaithilingam2022expectation, liang2024large, davila2024industry, marginean2019sapfix}, but significant challenges remain~\cite{liang2024large, vaithilingam2022expectation, barke2023grounded, weisz2025examining, winter2022towards}. A key limitation is overreliance on test suites for validation: tests may miss the target defect, leaving correctness uncertain. Wang et al.~\cite{wang2025solved} show benchmark-approved patches can fail developer-written tests or introduce unintended behavior. Prior work finds developers may accept overfitting fixes~\cite{eladawy2024automated}, skip inspection for large patches~\cite{cambronero2019characterizing}, or refactor AI-generated code to maintain quality~\cite{licorish2025comparing}. Nakashima et al.~\cite{nakashima2026agenticprsrejectedcomparativestudy} categorize rejection modes of agent-submitted pull requests, and Robbes et al.~\cite{robbes2026promisesperilstimelyheuristics} propose heuristics for mining coding agent activity on GitHub. Many studies examine correctness of automatically generated patches~\cite{xiong2018identifying,yang2023large,xin2017identifying,qi2015analysis}; our work is complementary and focuses on agentic repair workflows. \textbf{Understanding Agent Behavior through Trajectories.} Prior work analyzes SWE-bench trajectories for efficiency, structure, and failure patterns. Bouzenia and Pradel~\cite{bouzenia2025understandingsoftwareengineeringagents} study length, token use, and thought/action alignment, noting architectures may be task-dependent. Liu et al.~\cite{liu2025processcentricanalysisagenticsoftware} and Majgaonkar et al.~\cite{majgaonkar2025understandingcodeagentbehaviour} analyze open-process agents but omit workflow-based baselines, limiting architecture comparisons. Liu et al.~\cite{liu2025empiricalstudyfailuresautomated} build a failure-mode taxonomy across repair phases, while Yin et al.~\cite{yin2025comprehensiveempiricalevaluationagent} study general-purpose agents on code tasks with little APR-specific depth. Our work complements these studies by comparing five APR agents spanning workflow-based and open-process designs, using a trajectory-based taxonomy to contrast behavior across APR phases (patching, testing, and tool use). \textbf{Agent vs. Developer Code.} Prior work evaluates LLM outputs and patch correctness~\cite{rondon2025evaluating} or localizes edit sites~\cite{chen2024evaluating}, and Wang et al.~\cite{wang2025solved} assess test-passing correctness. Milanese et al.~\cite{milanese2026humanagentversushumanpull} study testing practices in human vs.\ human--AI PRs, and Hora and Robbes~\cite{hora2026codingagentsgeneratingovermocked} analyze code commits to compare how often AI vs.\ humans write tests and use mocks. We instead analyze how agent patches relate to human fixes via clone detection and construct-level analysis. \textbf{Issue Reproduction.} Issue reproduction is a prerequisite for repair and a key signal for validating fixes~\cite{mundler2024swt}, yet reproduction success can be low (e.g., Otter reports 29\% on SWT-Bench-Lite~\cite{ahmed2024tdd}). Agents often use inference-time scaling: \textit{parallel} sampling (Agentless~\cite{xia2024agentless}) or \textit{sequential} refinement (Otter~\cite{ahmed2025otter}, SpecRover~\cite{ruan2024specrover}). Prior work has studied scaling for general code generation~\cite{li2025s}, but not specifically for test generation. LIBRO~\cite{kang2023large} studies prompting for reproduction tests, and SWT-Bench~\cite{mundler2024swt} modifies agent prompts but does not isolate the impact of localization information on test generation. We study reproduction-test strategies within an end-to-end patch repair framework and quantify the effect of localization.

\FloatBarrier
\section{Threats to Validity}
\textbf{Evaluation methodology:} We use strong coding LLM backbones (GPT-4o, Claude 3.5 Sonnet), keep the backbone fixed within experiments, and confirm key trends across models. For LLM-based evaluation (e.g., RQ3) we use Claude Sonnet 4. Our LLM-based clone detection aligns with human judgment (Cohen's $\kappa = 0.81$ on 50 samples with two independent annotators). To reduce dependence on any single metric, we triangulate with LLM clone detection, Levenshtein distance, and manual inspection; AST/tree-edit methods are less applicable because patches are diffs that may not parse as complete programs. Where automated labeling is ambiguous (e.g., classifying agent-run tests), we manually validate random samples for each agent--LLM pair. \textbf{Human annotation bias:} Manual analyses involve judgment; we mitigate this with team review of borderline cases, consensus labeling, and explicit resolution of disagreements. Annotators consisted of co-authors and external student annotators. Annotation was performed independently, with the same samples assigned to multiple annotators to avoid any single reviewer dominating the process. After independent annotation, observations were consolidated and disagreements were adjudicated by a third reviewer (a co-author), ensuring no single perspective determined the final taxonomy. All taxonomy nodes were cross-validated against each agent's published source code and documentation to ground the taxonomy in objective evidence rather than annotator interpretation. \textbf{Sampling bias:} Results use a subset of SWE-bench Verified; we use Verified-50~\cite{zainullina2025guided} (shown to be similar to the full set~\cite{verified50blog}) and stratify sampling across difficulty levels to reduce skew. \textbf{Distribution shift / mixed authorship:} We evaluate fixed-snapshot, human-authored repositories; real codebases evolve and increasingly mix human- and AI-authored code, potentially changing failure modes and repair strategies. Current benchmarks do not capture this, so we leave it to future work (e.g., controlled injections of AI-authored edits). \textbf{Taxonomy coverage:} We do not cover all possible decision pathways in the taxonomy, but instead study representative, high-impact workflows spanning dominant strategies. \textbf{Agent choice:} We study five high-performing, architecturally diverse APR agents; results may not generalize to all systems, but many newer agents share underlying components. Although agents evolve rapidly, the dominant architectural paradigms on the SWE-bench leaderboard remain stable, with gains driven primarily by stronger backbone LLMs rather than new designs. Many newer agents inherit the workflow-based and open-process principles we study, making our findings on architectural bottlenecks transferable. Tool orchestration and test generation remain key bottlenecks even as LLMs improve, and simpler agents may be even more limited in addressing these challenges. While the trend toward minimalist agents is largely enabled by foundation model reasoning, tool-augmented designs remain necessary for complex, previously-unseen problems. Our taxonomy and subtask-level results provide a concrete baseline for comparing future agents across this design spectrum. We scope certain analyses to agent subsets, consistently selecting Agentless and SWE-Agent as representatives of each paradigm for which both backbones are available: RQ1 manual patch analysis, RQ2.3 overfitting analysis, and RQ3 reproducer analysis all use this pair, with high reproducer-generation rates providing sufficient samples. In Table 3, we use only Agentless due to computational cost and its highest reproducer-generation rate. RQ4 excludes Agentless and ACR because their workflow implementations preclude LLM action selection (Section 7.1). We study SWE-Agent rather than mini-SWE-agent because our goal is component-level traceability analysis. SWE-Agent provides a more modular scaffolding, exposing intermediate steps (planning, tool use, trajectory execution) in greater detail, while minimalist agents abstract away these details, making fine-grained analysis difficult. SWE-Agent's richer tool-calling interface also enables more informative cross-agent comparisons (e.g., RQ4). Both achieve competitive performance on SWE-bench Verified, so this choice does not compromise our empirical findings. Extending to more agents is future work. \textbf{Contamination:} While contamination is possible, several observations suggest robustness: (1) agents trigger bugs only ~50\% of the time; if memorization were dominant, success rates would likely be higher; (2) because our main findings are comparative of agents rather than LLMs, memorization would likely benefit all agents similarly. We observe consistent architecture-dependent differences (e.g., resolution/overfitting rates), suggesting performance is driven primarily by agent design and intermediate reasoning signals rather than recall.

\section{Conclusions}
We present a trace-based analysis of APR agent decision-making workflows. No single agent dominates: workflow agents produce concise, developer-aligned patches on simpler tasks, while open-process agents better manage complex refactorings but often generate more verbose solutions (up to 40$\times$ larger). Architectural scaffolding strongly shapes whether agents mimic developer patterns or pursue novel strategies; on harder issues, all agents increasingly diverge from developer style and plateau beyond $\sim$40 tasks. Across architectures, validation is a major bottleneck. Agents frequently fail to produce reproduction tests (triggering bugs only half the time) and struggle to select relevant regression tests; even workflow agents overfit to self-generated reproducers, leading to downstream failures. We find selective inference scaling and bug localization context can mitigate these issues, while higher-quality test generation strengthens validation and reduces spurious patches. Tooling also constrains performance: agents largely operate with primitive commands rather than debuggers/analyzers, and specialized APIs can measurably shift behavior (e.g., MASAI's $2\times$ higher Python execution frequency). \textbf{Actionable Insights:} (1) \textit{Hybrid architectures:} combine structured localization with flexible exploration for complex, multi-file fixes. (2) \textit{Shift-left testing:} prioritize reproduction-test generation and improve fail-to-pass quality to reduce spurious patches. (3) \textit{Localization + selective scaling:} use localization context to raise trigger rates; scale inference for well-specified issues and avoid waste on underspecified ones. (4) \textit{Decoupled, LLM-based test selection:} surface candidate tests separately, and filter early (before patching) to improve relevance (Agentless shows $12\times$ improvement with this approach). (5) \textit{Richer tooling:} integrate debuggers/static analyzers/specialized APIs instead of relying on generic shell commands. (6) \textit{Ensembles:} combine complementary agents since scaffolding induces different strengths (concise localized fixes vs. complex refactoring). \textbf{Implications for Benchmarks and Self-Evolving Agents:} Benchmarks should measure reproduction-test quality, regression-test selection, and semantic correctness beyond test passing/report stratified results (e.g., by bug complexity, edit scope, domain) to expose design tradeoffs. They should also model distribution shift in evolving, mixed-authorship codebases. Because our findings primarily reflect scaffolding-induced behaviors rather than LLM capability, they likely transfer to newer agent systems (e.g., SWE-bench Pro~\cite{deng2025swebenchpro} uses SWE-Agent scaffolding; SWE-RL~\cite{wei2025swerl} builds on Agentless), and motivate self-evolving agents that learn when to invoke richer tooling, route tasks to specialized sub-agents, switch between structured and flexible modes, exploit localization context, scale inference, and avoid overfitting to self-generated reproducers.

\section*{Data-Availability Statement}
We provide open-source artifacts and support replication and independent verification of our contributions~\cite{replicationPackage}. The replication package includes instructions for replicating each experiment, raw agent trajectories (hosted on Hugging Face), analysis scripts, and all data referenced throughout the paper.
\section*{Acknowledgments}

We thank Anvith Pabba and Anik Dey for their contributions to understanding the AutoCodeRover workflow and early explorations of temperature-based patching, and for contributing to the refinement of the taxonomy used in this study, respectively.

This work was supported in part by NSF CNS-2247370, NSF CCF-2313055, IBM, and the NSF Graduate Research Fellowship Program (GRFP). Any opinions, findings, conclusions or recommendations expressed herein are those of the authors and do not necessarily reflect those of the US Government, NSF, or IBM.

\bibliographystyle{ACM-Reference-Format}
\bibliography{sample-sigconf} 

\lstset{
  basicstyle=\footnotesize\ttfamily,
  breaklines=true,
  breakatwhitespace=true,
  columns=fullflexible,
  keepspaces=true,
  frame=single,
  language=C,
  xleftmargin=0.05\textwidth,
  xrightmargin=0.05\textwidth
}

\tcbset{
  promptstyle/.style={
    breakable,         
    enhanced,           
    colback=gray!10,
    colframe=white,
    boxrule=0pt,
    arc=0pt,
    left=6pt,
    right=6pt,
    top=4pt,
    bottom=4pt
  }
}

% Appendix content moved to annotations_standalone.tex for separate compilation

\end{document}